\documentclass[twocolumn]{aastex631}
\usepackage{amsmath}
\usepackage{amssymb}
\usepackage{graphicx}
\usepackage{enumitem}
\setcitestyle{notesep={; }}

\shorttitle{X-ray Evolution of the Changing-Look AGN 1ES 1927+654}
\shortauthors{Masterson et al.}

\begin{document}

\title{Evolution of a Relativistic Outflow and X-ray Corona in the Extreme Changing-Look AGN 1ES~1927+654}

\author[0000-0003-4127-0739]{Megan Masterson}
\affiliation{MIT Kavli Institute for Astrophysics and Space Research,
Massachusetts Institute of Technology, 
Cambridge, MA 02139, USA}
	
\author[0000-0003-0172-0854]{Erin Kara}
\affiliation{MIT Kavli Institute for Astrophysics and Space Research,
Massachusetts Institute of Technology, 
Cambridge, MA 02139, USA}

\author[0000-0001-5231-2645]{Claudio Ricci}
\affiliation{N\'{u}cleo de Astronom\'{i}a de la Facultad de Ingenier\'{i}a, Universidad Diego Portales,
Av. Ej\'{e}rcito Libertador 441, 
Santiago, Chile}
\affiliation{Kavli Institute for Astronomy and Astrophysics, 
Peking University, 
Beijing 100871, People's Republic of China}

\author[0000-0003-3828-2448]{Javier A. Garc\'{i}a}
\affiliation{Cahill Center for Astronomy and Astrophysics,
California Institute of Technology,
Pasadena, CA 91125, USA}
\affiliation{Dr. Karl Remeis-Observatory and Erlangen Centre for Astroparticle Physics, Sternwartstr. 7, D-96049 Bamberg, Germany}

\author[0000-0002-9378-4072]{Andrew C. Fabian}
\affiliation{Institute of Astronomy, 
University of Cambridge, 
Madingley Road, Cambridge CB3 0HA, UK}

\author[0000-0003-2532-7379]{Ciro Pinto}
\affiliation{INAF - IASF Palermo, 
Via U. La Malfa 153, 
I-90146 Palermo, Italy}

\author{Peter Kosec}
\affiliation{MIT Kavli Institute for Astrophysics and Space Research,
Massachusetts Institute of Technology, 
Cambridge, MA 02139, USA}

\author[0000-0003-4815-0481]{Ronald A. Remillard}
\affiliation{MIT Kavli Institute for Astrophysics and Space Research,
Massachusetts Institute of Technology, 
Cambridge, MA 02139, USA}

\author[0000-0002-1661-4029]{Michael Loewenstein}
\affiliation{Astrophysics Science Division, 
NASA Goddard Space Flight Center, 8800 Greenbelt Road, 
Greenbelt, MD 20771, USA}
\affiliation{Department of Astronomy, 
University of Maryland, 
College Park, MD 20742, USA}

\author[0000-0002-3683-7297]{Benny Trakhtenbrot}
\affiliation{School of Physics and Astronomy, 
Tel Aviv University, 
Tel Aviv 69978, Israel}

\author[0000-0001-7090-4898]{Iair Arcavi}
\affiliation{School of Physics and Astronomy, 
Tel Aviv University, 
Tel Aviv 69978, Israel}
\affiliation{CIFAR Azrieli Global Scholars Program, 
CIFAR, 
Toronto, Canada}

\correspondingauthor{Megan Masterson}
\email{mmasters@mit.edu}

\begin{abstract}

1ES~1927+654 is a paradigm-defying AGN and one of the most peculiar X-ray nuclear transients. In early 2018, this well-known AGN underwent a changing-look event, in which broad optical emission lines appeared and the optical flux increased. Yet, by July 2018, the X-ray flux had dropped by over two orders of magnitude, indicating a dramatic change to the inner accretion flow. With three years of observations with \textit{NICER}, \textit{XMM-Newton}, and \textit{NuSTAR}, we present the X-ray evolution of 1ES~1927+654, which can be broken into three phases—(1) an early super-Eddington phase with rapid variability in X-ray luminosity and spectral parameters, (2) a stable super-Eddington phase at the peak X-ray luminosity, and (3) a steady decline back to the pre-outburst luminosity and spectral parameters. For the first time, we witnessed the formation of the X-ray corona, as the X-ray spectrum transitioned from thermally-dominated to primarily Comptonized. We also track the evolution of the prominent, broad 1~keV feature in the early X-ray spectra and show that this feature can be modeled with blueshifted reflection ($z=-0.33$) from a single-temperature blackbody irradiating spectrum using \texttt{xillverTDE}, a new flavor of the \texttt{xillver} models. Thus, we propose that the 1~keV feature could arise from reflected emission off the base of an optically thick outflow from a geometrically thick, super-Eddington inner accretion flow, connecting the inner accretion flow with outflows launched during extreme accretion events (e.g. tidal disruption events). Lastly, we compare 1ES~1927+654 to other nuclear transients and discuss applications of \texttt{xillverTDE} to super-Eddington accretors.

\end{abstract}



\section{Introduction} \label{sec:intro}

Active galactic nuclei (AGN) are characterized by accretion onto the supermassive black hole at the center of the host galaxy. The accretion process is intrinsically turbulent, leading to stochastic variability observed across the electromagnetic spectrum. Occasionally, more drastic changes in AGN emission are observed in objects called ``changing-look AGN" (CLAGN), in which the source rapidly transitions from one spectral type to another through the appearance or disappearance of broad optical emission lines \citep[e.g.][]{Shappee2014,LaMassa2015,MacLeod2016,Ruan2016,MacLeod2019,Yang2018,Hon2020}. The appearance or disappearance of optical broad lines is also commonly accompanied by an increase or decrease in the UV/X-ray continuum flux, respectively. These continuum changes, along with significant changes in the IR fluxes, low levels of optical polarization, and complex multi-wavelength variability suggest that these changing-look events are driven by an intrinsic change in mass accretion rate \citep[e.g.][]{Sheng2017,Mathur2018,Stern2018,Hutsemekers2019}, rather than being a transient obscuration effect. However, the mechanisms which induce these rapid changes in mass accretion rate are still not well understood, with current theories including binary supermassive black holes \citep{Wang2020b}, state transitions similar to X-ray binaries \citep{Noda2018,Ruan2019,Ai2020}, magnetically elevated accretion in thick disks \citep{Dexter2019}, and instabilities in the accretion disk \citep{Ross2018,Sniegowska2019}. 

1ES 1927+654 is a well-known nearby AGN ($z = 0.019422$) that was first discovered with the \textit{Einstein} satellite \citep{Elvis1992,Perlman1996} and has recently gone through an extreme changing-look event. Early X-ray observations of 1ES~1927+654 taken decades before the changing-look event showed little evidence for obscuration by dust along the line of sight with $N_H \approx 10^{21}$ cm$^{-2}$, yet no broad emission lines were detected in optical observations, even in polarized emission \citep{Boller2003,Tran2011}. This made 1ES~1927+654 part of the ``True Type 2" class of AGN, which seem to defy simple unified AGN models \citep[e.g.][]{Antonucci1993,Urry1995}, although recent studies have found that the broad lines seemingly missing in some True Type 2 AGN are simply too broad or too faint to be easily detected \citep[e.g.][]{Bianchi2019}. This classification prompted further X-ray observations of 1ES~1927+654, including observations with \textit{ROSAT} in the 1990s, \textit{Chandra} in 2001, and \textit{Suzaku} and \textit{XMM-Newton} in 2011. All of these observations showed that the X-ray spectrum was dominated by a steep power law ($\Gamma \approx 2.4$) and a soft excess with no evident narrow iron~K line \citep{Boller2003,Gallo2013}.

This unique source became even more interesting when the All-Sky Automated Survey for Supernovae \citep[ASAS-SN;][]{Shappee2014} reported a significant and rapid optical brightening of 1ES~1927+654 of at least two magnitudes in the V band on 2018 March 3 \citep[ASASSN-18el/AT2018zf;][]{Nicholls2018}. Archival data from the Asteroid Terrestrial-impact Last Alert System \citep[ATLAS;][]{Tonry2018} revealed that the outburst actually began around 2017 December 23. The discovery of this extreme optical transient in an existing AGN defied what would be expected for a typical AGN flare and prompted follow-up observations across the electromagnetic spectrum. Optical spectroscopy immediately following the detection of the optical outburst revealed a featureless, blue continuum, reminiscent of a quasar optical spectrum. Broad H$\alpha$ and H$\beta$ emission lines then appeared within a few months of the initial outburst detection, making 1ES~1927+654 a CLAGN and the first object to have been observed undergoing this transition in real time, over timescales of months \citep{Trakhtenbrot2019}. The change in optical state motivated high cadence X-ray monitoring of 1ES~1927+654 with \textit{XMM-Newton}, \textit{NuSTAR}, \textit{NICER}, and \textit{Swift} beginning in late May 2018. 

The first follow-up X-ray observations revealed that the X-ray flux was near its pre-outburst level in 2011, but the spectrum was starkly different from its pre-outburst spectrum, dominated by an extremely soft, thermal component with very little emission above 3~keV \citep{Ricci2020}. The X-ray flux then dropped by over an order magnitude, followed by an increase in X-ray flux of approximately four orders of magnitude to the Eddington limit of a $10^6$ M$_\odot$ black hole, over the span of a few hundred days \citep{Ricci2020}. One possible explanation for this extreme variability and change in the X-ray spectrum is that a tidal disruption event (TDE) occurred in 1ES~1927+654, causing the depletion of the inner accretion disk and cutting off the energy supply to the corona that produces the hard X-ray emission in AGN. In addition, the early X-ray data revealed a broad emission-like feature at 1~keV, which is very prominent, but not well understood \citep{Ricci2021}. These fascinating and unique early X-ray observations motivated extensive X-ray monitoring, spanning more than three years with seven simultaneous \textit{XMM-Newton}/\textit{NuSTAR} observations and more than 500 \textit{NICER} observations. To date, 1ES~1927+654 has been observed for more than 1 Ms with \textit{NICER} and is currently the most observed AGN in the \textit{NICER} archive.

In this work, we present the X-ray evolution of the full outburst in 1ES~1927+654, extending from the initial X-ray observations taken in May 2018 to when it had decayed down to its pre-outburst level in June 2021. The first half of this monitoring campaign was first presented in \citet{Ricci2020,Ricci2021}, and in this work, we present new observations, including over 200 new \textit{NICER} observations four new simultaneous \textit{XMM-Newton} and \textit{NuSTAR} observations, from the second half of the outburst, as outlined in Table \ref{tab:obs}. In Section \ref{sec:obs}, we discuss the data used in this analysis and the data reduction processes. We present the X-ray spectral evolution of the source in Section \ref{sec:evolution}. In Section \ref{sec:physmod}, we shift our focus to a physically-motivated reflection model to explain the prominent broad 1~keV line in the early X-ray observations. Lastly, we discuss the impact of our findings on our understanding of 1ES~1927+654 and other TDEs, CLAGN, and nuclear transients in Section \ref{sec:discussion} and summarize our results in Section \ref{sec:conclusion}. Throughout this paper we adopt a standard $\Lambda$CDM cosmology with $H_0 = 70$ km s$^{-1}$ Mpc$^{-1}$, $\Omega_\Lambda=0.73$, and $\Omega_M = 0.27$. All quoted errors are 90\% confidence ($\Delta \chi^2 = 2.706$ for one parameter of interest).


\section{Observations \& Data Reduction} \label{sec:obs}

1ES~1927+654 has been extensively monitored since its late 2017 optical outburst, with a particular focus on the dramatic evolution in the X-ray band. Together \textit{XMM-Newton} \citep{Jansen2001} and \textit{NuSTAR} \citep{Harrison2013} have performed seven simultaneous pointed observations of 1ES~1927+654 with roughly 4-6 months cadence. Details of the simultaneous \textit{XMM-Newton} and \textit{NuSTAR} observations are given in Table \ref{tab:obs}. Given how soft and bright 1ES~1927+654 has been, \textit{NICER} \citep{Gendreau2012,Arzoumanian2014} has observed 1ES~1927+654 with a much higher cadence of approximately one observation every two days. 

Here we present the X-ray observations and data reduction for the follow-up campaign of 1ES~1927+654 up to June 2021, including the first half of the monitoring campaign presented in \citet{Ricci2020,Ricci2021} and the latter half of the monitoring campaign presented here for the first time. All proceeding X-ray spectral analysis was performed using XSPEC version 12.11.1 \citep{Arnaud1996} with $\chi^2$ fit statistics.

\begin{deluxetable*}{c c c c c c c c}

	\caption{\textit{XMM-Newton} and \textit{NuSTAR} Observation Information} \label{tab:obs}
	
    \tablehead{\colhead{Epoch} & \colhead{Date} & \colhead{Telescope} & \colhead{ObsID} & \colhead{Exposure} & \colhead{First Presented} & \colhead{Soft Count Rate}\tablenotemark{$\dagger$} & \colhead{Hard Count Rate}\tablenotemark{$\dagger$} \\
    \colhead{} & \colhead{} & \colhead{} & \colhead{} & \colhead{(ksec)} & \colhead{} & \colhead{(cts s$^{-1}$)} & \colhead{(cts s$^{-1}$)}}

	\startdata
	1 & 2018-06-05 & \textit{XMM} & 0830191101 & 46.4 & \citet{Ricci2020,Ricci2021} & $7.76 \pm 0.02$ & $0.0077 \pm 0.0005$ \\
	&  & \textit{NuSTAR} & 90401625002 & 45.9 & & -- & -- \\
	2 & 2018-12-12 & \textit{XMM} & 0831790301 & 59.3 & \citet{Ricci2020,Ricci2021} & $49.06 \pm 0.04$ & $0.719 \pm 0.004$ \\
	&  & \textit{NuSTAR} & 90401641002 & 64.7 & & $0.0336 \pm 0.0007$ & $0.0080 \pm 0.0004$ \\
	3 & 2019-05-06 & \textit{XMM} &0843270101 & 52.0 & \citet{Ricci2020,Ricci2021} & $52.34 \pm 0.04$ & $1.137 \pm 0.006$ \\
	&  & \textit{NuSTAR} & 90501618002 & 58.2 & & $0.066 \pm 0.001$ & $0.0254 \pm 0.0007$ \\
	4 & 2019-11-02 & \textit{XMM} & 0843270201 & 53.5 & This work & $74.74 \pm 0.05$ & $2.216 \pm 0.008$ \\
	&  & \textit{NuSTAR} & 60502034002 & 50.7 & & $0.124 \pm 0.002$ & $0.046 \pm 0.001$ \\
	5 & 2020-05-03 & \textit{XMM} & 0863230101 & 47.0 & This work & $40.83 \pm 0.04$ & $1.626 \pm 0.007$ \\
	&  & \textit{NuSTAR} & 60502034004 & 45.8 & & $0.133 \pm 0.002$ & $0.052 \pm 0.001$ \\
	6 & 2020-09-16 & \textit{XMM} & 0863230201 & 49.8 & This work & $14.64 \pm 0.02$ & $1.046 \pm 0.006$ \\
	&  & \textit{NuSTAR} & 60602003002 & 36.1 & & $0.132 \pm 0.002$ & $0.137 \pm 0.002$ \\
	7 & 2021-01-12 & \textit{XMM} & 0863230301 & 48.1 & This work & $2.21 \pm 0.01$ & $0.304 \pm 0.003$ \\
	&  & \textit{NuSTAR} & 60602003004 & 48.8 & & $0.048 \pm 0.001$ & $0.073 \pm 0.001$ \\
    \enddata
    
    \tablenotetext{\dagger}{Count rates are averaged over the entire observation, and uncertainties reflect the Poisson noise. For \textit{XMM-Newton}, the soft count rate is measured in the 0.3-2 keV range, and the hard count rate is measured in the 2-10 keV range. For \textit{NuSTAR}, the soft count rate is measured in the 3-5 keV range, and the hard count rate is measured in the 5-20 keV range.}
    
\end{deluxetable*}

\subsection{XMM-Newton} \label{subsec:xmm}

1ES~1927+654 was detected with each instrument on board \textit{XMM-Newton} during all seven observations. For the purposes of this work, we focus on data from the EPIC-pn camera, which has a higher effective area and is less sensitive to issues of pile-up than EPIC-MOS. Pile-up was particularly prevalent in the high-luminosity observations of 1ES~1927+654, motivating the use of only the EPIC-pn data. Given extensive observations with \textit{NICER} as well as the EPIC-pn data, the results of this work are not statistics-limited and EPIC-MOS data was not necessary. Likewise, we do not perform extensive modeling with the RGS data, as \citet{Ricci2021} found that although there was a weak ionized absorber in the soft spectrum, the 1~keV line that is the focus of this paper is still broad in the RGS spectrum. Thus, the main features studied in this work (the temperature evolution of the blackbody, the power law evolution, and the 1~keV feature) are all broad spectral components that are best constrained with broad-band modeling of EPIC-pn data. We do find that the best-fitting physical model for the Epoch 1 EPIC-pn spectrum (discussed in Section \ref{subsec:june2018}) provides an statistically acceptable fit to the RGS spectrum, even without including the weak ionized absorber found in \citet{Ricci2021}. Similarly, multi-wavelength analysis using the optical and UV data from the OM instrument on \textit{XMM-Newton} will be presented in a forthcoming paper focused on broad-band modeling of the spectral energy distribution (SED) of 1ES 1927+654 (Li et al. in prep.).

We reduced the EPIC-pn data from each \textit{XMM-Newton} observation using the \textit{XMM-Newton} Science Analysis System (SAS; version 18.0.0) with latest calibration files. We followed standard data reduction procedures for EPIC-pn, including running \texttt{epproc} to process the raw data and make calibrated event lists. We made standard cuts on high energy count rates to avoid periods of background flaring, excluding any time where the 10-12~keV count rate was above 0.4 cts s$^{-1}$. As 1ES~1927+654 was extremely bright and soft during most observations, we used annular extraction regions for all epochs to minimize the effects of pile-up. In Epochs 1, 6, and 7, pile-up was present, but not as significant as in the other observations, so we adopted an inner (outer) extraction radius of 6" (40") for these data. For Epochs 2-5, the effects of pile-up were significant given the source brightness and hence, we used an inner (outer) extraction radius of 15" (40"). In addition, where possible, we compared the \textit{XMM-Newton} spectra to the \textit{NICER} spectra, which do not suffer from the effects of pile-up, and found good agreement in spectral shape between the two using angular extraction regions for the \textit{XMM-Newton} data. We extracted a background spectrum from an off-source circular region on the same CCD chip with a radius of 35". We utilized all single and double events (PATTERN $\leq$ 4) when extracting spectra. Redistribution matrix files and ancillary response files for each observation were created using \texttt{rmfgen} and \texttt{arfgen}, respectively. Finally, the spectra were grouped to have a minimum of 25 counts per bin.

\subsection{NuSTAR} \label{subsec:NuSTAR}

1ES~1927+654 was simultaneously observed with \textit{NuSTAR} during all seven \textit{XMM-Newton} observations, but was undetected with \textit{NuSTAR} in the first observation \citep{Ricci2021}. For the remaining six observations, we reduced the \textit{NuSTAR} data using \textit{NuSTAR} Data Analysis Software (NuSTARDAS; version 2.0.0 in HEASoft version 6.28) with calibration files from \textit{NuSTAR} CALDB v20210104. We followed standard data reduction procedures for \textit{NuSTAR} data, processing the data with \texttt{nupipeline} and extracting spectra for both the FPMA and FPMB modules with \texttt{nuproducts}. Spectra were extracted from circular regions around the source with a radius of 50". Background spectra were extracted from a circular off-source region with a radius of 80". Given the softness of the source, pile-up was not evident in the \textit{NuSTAR} data. The spectra were again grouped to have a minimum of 25 counts per bin.

\subsection{NICER} \label{subsec:NICER}

High cadence X-ray observations of 1ES~1927+654 were taken with \textit{NICER}, with typical time between observations of a few hours to a few days. \textit{NICER} monitoring of 1ES~1927+654 began on 22 May 2018, and in this work, we report the results of all observations taken up to 21 June 2021. We reduced \textit{NICER} observations using tools in the NICERDAS suite (HEASoft version 6.28) with the 2020 gain. We exclude data from focal plane detector modules 14 and 34, which are known to be excessively noisy, and use an appropriately weighted ARF and RMF file for the remaining 50 detectors. 

Unlike with \textit{XMM-Newton} and \textit{NuSTAR} observations where an X-ray background can be measured using off-source CCD pixels, \textit{NICER} background must be estimated using spectral parameters as there is no off-source region. We implement the 3C50 background model for our \textit{NICER} observations \citep{Remillard2022}, and filter on the background-subtracted spectra. Namely, we expect that below 0.2~keV and above 13~keV, the background-subtracted count rates should be sufficiently close to zero, since the detector has negligible effective area outside of 0.2-13~keV. Any strong deviation from net zero flux in these ranges indicates an issue with the background modeling. Thus, we exclude any spectra where the absolute value of the count rate in the 13-15~keV range is $> 0.1$ cts s$^{-1}$ or the absolute value of the count rate below $0.2$~keV is $> 10$ cts s$^{-1}$ from our fitting \citep[level 2 filtering as described in][]{Remillard2022}. All of the spectra were grouped on a per-ObsID basis and to a minimum of 25 counts per bin. 


\section{Spectral Evolution} \label{sec:evolution}

\begin{figure*}[t!]
    \centering
    \includegraphics[width=17.5cm]{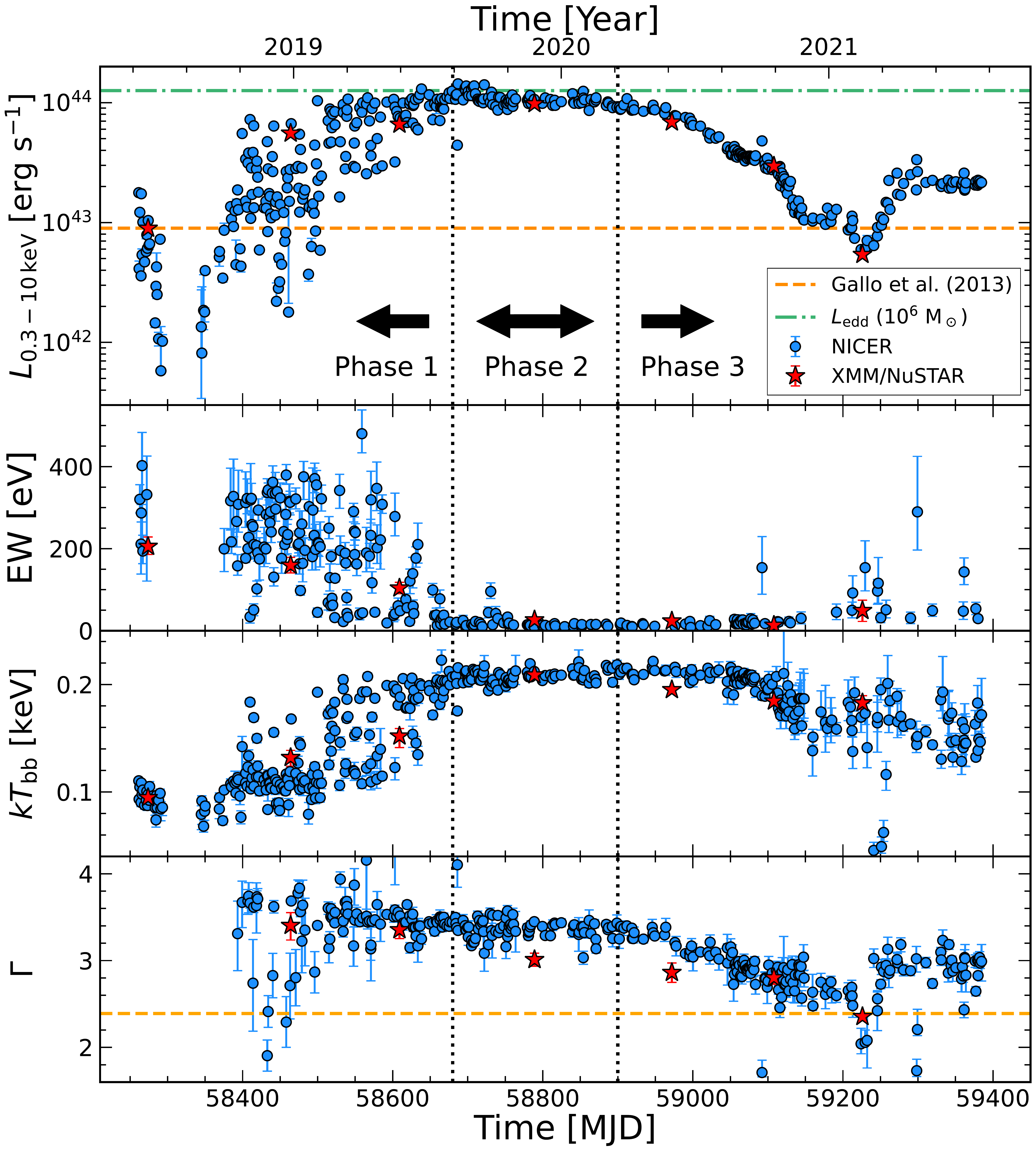}
    \caption{Evolution of various aspects of the X-ray spectrum of 1ES~1927+654 from May 2018 through June 2021. In all panels, blue circles correspond to a single \textit{NICER} ObsID and red stars are measurements from simultaneous \textit{XMM-Newton}/\textit{NuSTAR} observations. Where shown, the orange dashed line corresponds to the pre-outburst level reported in \citet{Gallo2013}. The vertical dashed black lines highlight a unique three-phase evolution, which were identified based on properties of the X-ray light curve and changes in the X-ray spectral properties. \textit{Top panel:} X-ray light curve of 1ES~1927+654 showing the 0.3-10~keV luminosity. The green dot-dashed line shows the Eddington limit for a $10^6$ M$_\odot$ black hole, highlighting the plateau in the X-ray light curve. \textit{Second panel:} Equivalent width of the 1~keV Gaussian line. Only fits with $< 50\%$ uncertainty on the equivalent width measurement and with Gaussian line significance of $> 99.99\%$ are shown here. \textit{Third panel:} Temperature of the blackbody spectral component. \textit{Bottom panel:} Evolution of the photon index of the power law spectral component. We exclude the \textit{NICER} data points where $\Gamma$ was fixed at 3 or where the relative uncertainty on $\Gamma$ was greater than 20\% (see Section \ref{subsec:phenom_model}).}
    \label{fig:lumin}
\end{figure*}

In Figure \ref{fig:lumin}, we show the evolution of the X-ray properties of 1ES~1927+654 over the course of the last three years with \textit{NICER}, \textit{XMM-Newton}, and \textit{NuSTAR} observations. The long-term light curve of 1ES~1927+654 (with X-ray luminosities measured from the modeling in Section \ref{subsec:phenom_model}), shown in the top panel, is unlike any accreting source we have ever observed before. The high-cadence \textit{NICER} observations show rapid variability on the X-ray rise, contrasted with an extremely smooth decline in X-ray flux. The orange dashed line indicates the pre-outburst X-ray luminosity reported by \citet{Gallo2013} based on the \textit{XMM-Newton} and \textit{Suzaku} observations from 2011. At late times, the X-ray luminosity seems to be trending back toward its pre-outburst state and the X-ray spectrum is remarkably similar to the pre-outburst X-ray spectrum, which we discuss further in Section \ref{subsec:comp2arch}. The 0.3-10 keV X-ray luminosity of 1ES~1927+654 appears to reach its maximum around the Eddington limit for a $10^6$ M$_\odot$ black hole as shown in the green dot-dashed line in the top panel of Figure \ref{fig:lumin}. As the X-ray spectrum of 1ES~1927+654 is unlike any AGN ever observed before, we cannot apply a simple bolometric correction to estimate the bolometric luminosity and accretion rate. However, multi-wavelength observations with high optical/UV luminosity at early times suggest an extended super-Eddington phase until near the point at which the X-ray luminosity decreases (Li et al. in prep).

In Figure \ref{fig:lumin}, we also show three different evolutionary phases separated by vertical black dotted lines. These phases are identified based on the spectral properties and evolution of the light curve. The first phase corresponds to the early portion of the light curve, from first X-ray observations to mid-2019 (MJD $<$ 58680), and is dominated by rapid X-ray variability and a weak power law component. We chose the second phase to start where the X-ray light curve became very stable with little variability. This phase extends until an MJD of 58900, at which point the X-ray luminosity starts to drop and the photon index of the power law begins to decrease.

\subsection{Phenomenological Modeling} \label{subsec:phenom_model}

\subsubsection{Phenomenological Modeling with NICER}

To assess the evolution of the X-ray spectral shape, we applied a simple phenomenological model to all of the NICER data, including galactic and source absorption, a power law, and a blackbody (i.e. \texttt{tbabs*ztbabs*(zbb+zpower)} in XSPEC notation). In addition to these simple continuum components, the early X-ray observations of 1ES~1927+654 also reveal a prominent emission-like excess around 1~keV. We explore a physically-motivated reflection model of this feature in Section \ref{sec:physmod}, but for the purposes of tracking the line evolution, we include a Gaussian emission line model in our simple phenomenological model. Thus, our overall phenomenological model is \texttt{tbabs*ztbabs*(zbb+zpower+zgauss)} in XSPEC notation. The line is only included in our modeling when it is significant at the 99.99\% level, using the F-test as an approximate measure of statistical significance. As discussed further in Section \ref{subsec:disapp_1keV}, we only included the line up until the end of Phase 1 (up to an MJD of 58680) when the source has reached its peak X-ray luminosity, at which point the line has either disappeared or is indistinguishable from a broad continuum component. Thus, we fit each NICER observation with one of two models, either \texttt{tbabs*ztbabs*(zbb+zpower)} or \texttt{tbabs*ztbabs*(zbb+zpower+zgauss)}, depending on the above criteria for the inclusion of a Gaussian emission line.

When fitting, we adopted abundances from \citet{Wilms2000} and cross-sections from \citet{Verner1996}. The redshift of each component was fixed at the source redshift of $z = 0.019422$. Galactic absorption from the \texttt{tbabs} model was fixed at $N_H = 6.42 \times 10^{20}$ cm$^{-2}$ \citep{HI4PICollaboration2016}. The fit parameters for the base continuum model were the column density, blackbody temperature, blackbody normalization, photon index and power law normalization in each observation. When the Gaussian line was included in the model, the line energy, width, and normalization were all free in spectral fitting, but we bounded the Gaussian line energy to be between 0.8 and 1.2~keV and the Gaussian line width to be $< 0.3$~keV. These choices were made to ensure that the Gaussian line was picking up the 1~keV feature and not blurring into some additional continuum component.

When fitting \textit{NICER} data, we only considered energies where the source flux is higher than the background. This included data from 0.3~keV up to an upper bound $E_\mathrm{upper}$ which varied from less than 1~keV up to 5~keV over the observation period. For $E_\mathrm{upper} < 2.2$~keV, we found that the photon index was poorly constrained with a fractional uncertainty greater than 20\%. Hence, we chose to fix the photon index at $\Gamma = 3$ \citep[as in][]{Ricci2021} for any \textit{NICER} observations where $E_\mathrm{upper} < 2.2$~keV. We also tested a cutoff power law with \textit{NICER} observations, but found that there was a significant degeneracy between the photon index and the cutoff energy given the low values of $E_\mathrm{upper}$. 

Given the large number of \textit{NICER} observations, we automated this fitting procedure and thus needed to apply some simple cuts to ensure that we are only including converged fits. We made a simple $\chi^2_\nu$ cut, keeping only the fits that have $\chi^2_\nu < 2$. This may have removed some observations with additional spectral complexity, but our goal with the phenomenological modeling was to provide a cohesive picture for the X-ray continuum evolution in 1ES 1927+654. In Section \ref{sec:physmod}, we perform more detailed physical modeling of the \textit{XMM-Newton}/\textit{NuSTAR} spectra, where we address these additional spectral complexities. Additionally, we noticed a few observations with poor background estimation, which showed up in the fitting procedure as artificially low photon indices in the \texttt{zpower} model (pegged at the minimum allowed $\Gamma$ of 1.4), and also excluded those observations from our results. These observations were sparsely and randomly spaced throughout the light curve, indicating that this likely was not an intrinsic change in the source properties. Finally, we only kept fits with sufficient constraints on the temperature of the blackbody and the photon index of the power law component, implementing a cut at 20\% fractional uncertainty for both of these parameters. These conservative cuts left us with 438 \textit{NICER} ObsIDs out of a total 495 ObsIDs considered, totalling more than 1 Ms of observation time presented in this analysis. In Appendix \ref{sec:app1}, we show the phenomenological modeling for three \textit{NICER} observations, which were taken close to simultaneous \textit{XMM-Newton}/\textit{NuSTAR} observations and are representative of the three evolutionary phases highlighted in Figure \ref{fig:lumin}.

\subsubsection{Phenomenological Modeling with XMM-Newton \& NuSTAR}

We followed a similar procedure for each simultaneous \textit{XMM-Newton}/\textit{NuSTAR} observation, jointly fitting the two spectra for each observation period with a simple phenomenological model. Likewise, we followed the same procedure as with the \textit{NICER} analysis for deciding whether or not to include a Gaussian emission line in the modeling. For the \textit{XMM-Newton} data, we only consider data between 0.3~keV and wherever the spectrum becomes background dominated, which varied between 3~keV (Epoch 1) and 10~keV (Epoch 6). Similarly, for the six \textit{NuSTAR} observations in which the source was detected, we considered data between 3~keV and wherever the background dominated, which varied between 8~keV (Epoch 2) and 20~keV (Epoch 7). The addition of the \textit{NuSTAR} data up to a higher energy allowed us to break the degeneracy between the photon index and cutoff energy in a cutoff power law model and required a cutoff in the power law distribution to most accurately describe the spectrum. Hence, when jointly fitting the \textit{XMM-Newton}/\textit{NuSTAR} data, we used a cutoff power law instead of a simple power law (i.e. \texttt{tbabs*ztbabs*(zbb+zcutoffpl)} or \texttt{tbabs*ztbabs*(zbb+zcutoffpl+zgauss)} in XSPEC notation). We fixed the cutoff energy at 300~keV when the cutoff energy is poorly constrained, which occurred in Epoch 1 when the source was undetected with \textit{NuSTAR} and in Epoch 7 when the cutoff energy was outside of the \textit{NuSTAR} bandpass. We note that fitting with a cutoff power law instead of a simple power law gives slightly lower photon indices compared to the \textit{NICER} values (as seen in the bottom panel of Figure \ref{fig:lumin}), which is in line with what we would expect to see when neglecting a high energy cutoff in the \textit{NICER} data. We show results of this phenomenological modeling to three of the \textit{XMM-Newton}/\textit{NuSTAR} observations in Appendix \ref{sec:app1}, with one observation from each evolutionary phase which are used later in Section \ref{sec:physmod} to do physically-motivated spectral modeling.

\subsection{Disappearance of the 1~keV Line} \label{subsec:disapp_1keV}

\begin{figure}[t!]
    \centering
    \includegraphics[width=8.6cm]{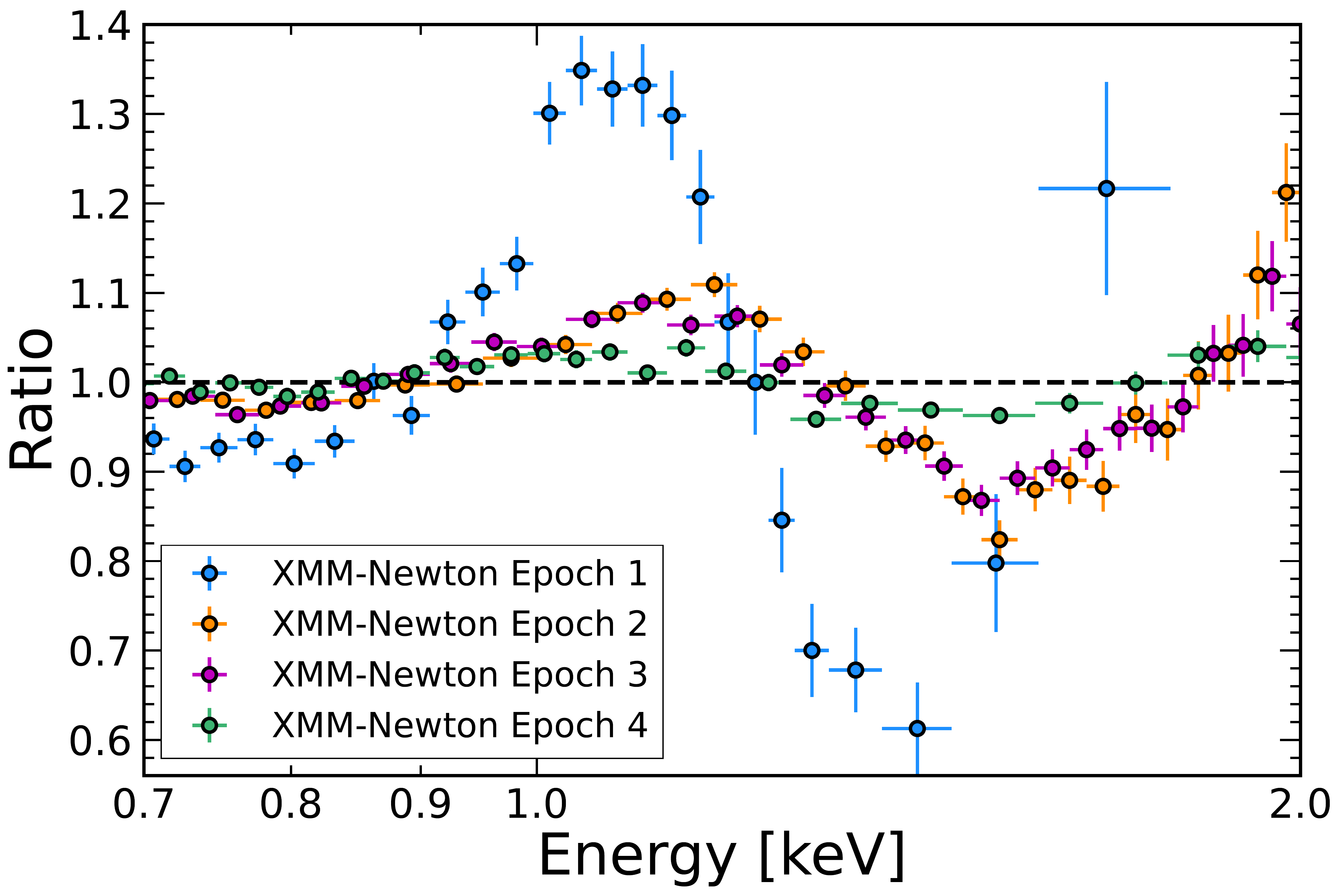}
    \caption{Ratio of data fit to a simple continuum model with a cutoff power law and blackbody for the first four \textit{XMM-Newton} observations. The 1~keV feature is evident in the first three observations, but is not as prominent in the fourth observation. The first observation shows the strongest 1~keV feature. Data have been binned for visual purposes only.}
    \label{fig:1keV_eqw}
\end{figure}

In Figure \ref{fig:1keV_eqw}, we show a ratio plot of the first four \textit{XMM-Newton} observations relative to the cutoff power law and blackbody phenomenological model. The base continuum model indicates an emission line at 1~keV and/or an absorption line between 1 and 2~keV, but the absorption residual is removed when including a 1~keV emission line in the spectral model. There is a clear trend in the strength of the 1~keV feature with time; the feature is strongest in the Epoch 1 \textit{XMM-Newton} observation and decreases in strength during the next two epochs. In the Epoch 4 spectrum, no strong 1~keV feature is present, although there are possibly additional broad continuum features that may arise from super-Eddington accretion during the peak X-ray luminosity phase. 

To assess the evolution of the 1~keV line, we looked at the equivalent width of the Gaussian model as a function of time. For this analysis, we used the same phenomenological modeling previously discussed and froze the Gaussian line at 1~keV to avoid a degeneracy between a weak line around 1~keV and an extremely broad line at the extremes of the model. This is shown for both \textit{NICER} and \textit{XMM-Newton} data in blue circles and red stars, respectively, in the second panel of Figure \ref{fig:lumin}. There is a large amount of scatter in the \textit{NICER} data at early times, which can be partially attributed to the large variability in luminosity of the \textit{NICER} observations as the observations with higher luminosity have the lower equivalent widths. 

Both sets of observations show the same general trends. Namely, the 1~keV line shows a very clear transition from large equivalent widths to extremely low equivalent widths ($\lesssim 20$~eV) around an MJD of 58680. This corresponds roughly to when the X-ray luminosity in the top panel of Figure \ref{fig:lumin} reaches its plateau at around $L_{0.3-10\, \mathrm{keV}} \approx 1.2 \times 10^{44}$ erg s$^{-1}$. To ensure that the decrease in equivalent width was not just the result of an increased continuum, we also checked the line flux, which also dropped slightly at high luminosities, indicating that both factors play a role in observing such small equivalent widths at peak X-ray luminosity. After the plateau, the X-ray luminosity drops back to its pre-outburst level (similar to that of Epoch 1), yet the 1~keV line does not reappear. Hence, for the remainder of the phenomenological modeling presented, we exclude the Gaussian component from the model beyond an MJD of 58680, which is shown as a vertical black dashed line (transition from Phase 1 to Phase 2) in Figure \ref{fig:lumin}. This boundary is near where the analysis performed by \citet{Ricci2021} ends, so we refer the reader to that work for a detailed description of the line evolution before it disappeared. In Section \ref{sec:physmod}, we present a physically-motivated reflection model to explain the 1~keV feature and discuss that how the evolution could be the result of suppression due to over-ionization effects. 

\subsection{Temperature Evolution} \label{subsec:temp_evol}

\begin{figure}[t!]
    \centering
    \includegraphics[width=8.6cm]{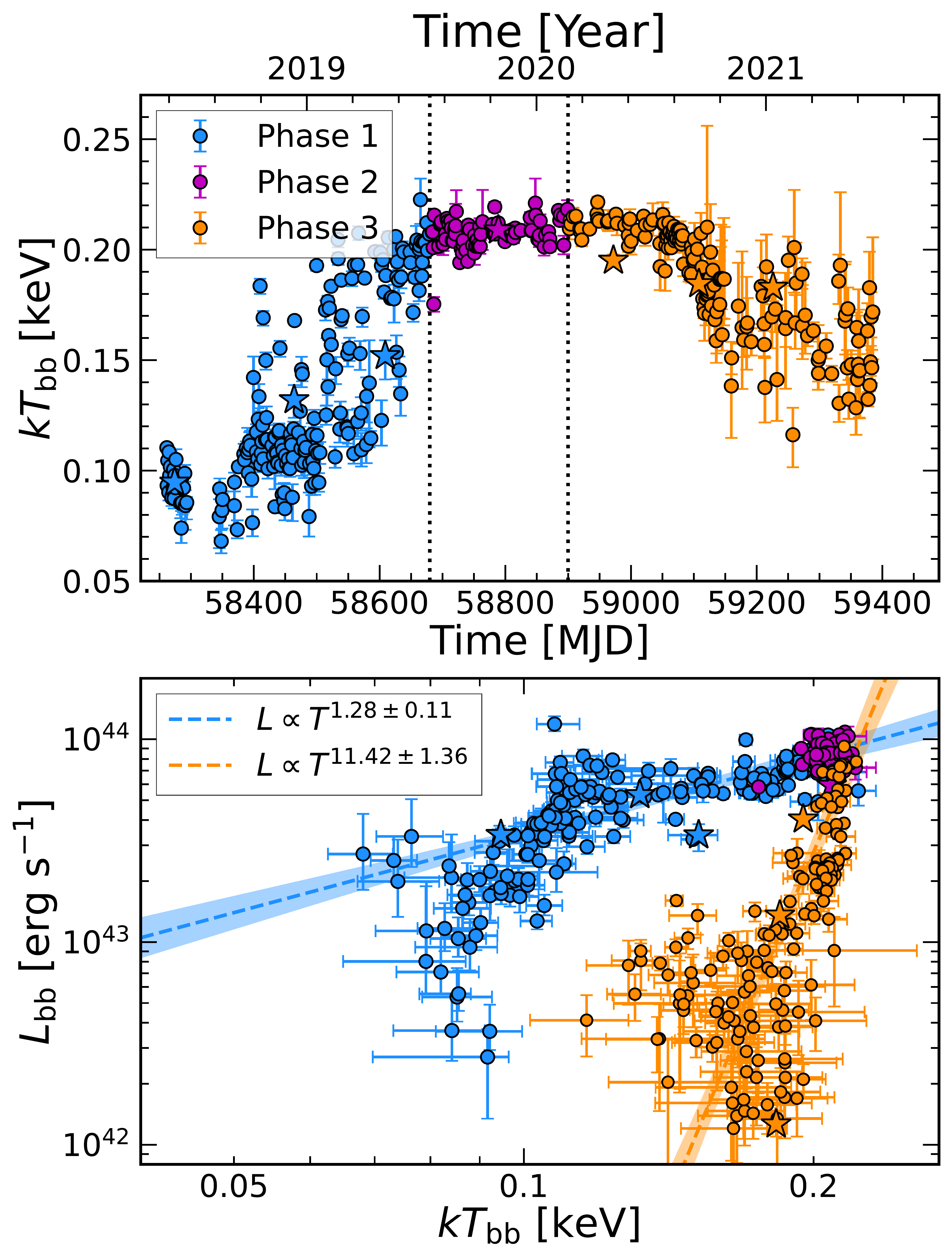}
    \caption{\textit{Top}: Temperature of the blackbody as a function of time. The \textit{NICER} data points are shown as circles and the \textit{XMM-Newton}/\textit{NuSTAR} data points are shown as stars. The data points are colored based on the phase from Figure \ref{fig:lumin}, with blue, magenta, and orange corresponding to Phase 1, 2, and 3 respectively. The phase boundaries from Figure \ref{fig:lumin} overplotted as black, vertical dotted lines. \textit{Bottom}: Luminosity of the blackbody as a function of the temperature. Again, the \textit{NICER} data points are shown as circles and the \textit{XMM-Newton}/\textit{NuSTAR} data points are shown as stars, which are again colored by the evolutionary phase. Dashed lines show the best fit to $L \propto T^n$ for Phases 1 and 3, with the shaded regions corresponding to the 90\% confidence intervals. The legend gives the exponent and uncertainty for each fit.}
    \label{fig:tempevol}
\end{figure}

Shortly after the outburst began, the X-ray spectrum of 1ES~1927+654 was well-characterized by a relatively hot blackbody with a temperature of $kT_\mathrm{bb} \approx$ 0.1~keV, which increased up to close to 0.2~keV as the X-ray luminosity increased \citep{Ricci2021}. However, standard thin accretion disks around supermassive black holes are expected to be around $kT_\mathrm{bb} \lesssim 0.05$~keV. Thus, this hotter blackbody emission could be indicative of emission from a hot, super-Eddington inner accretion flow. The top panel of Figure \ref{fig:tempevol} shows the evolution of the blackbody temperature as a function of time over the entire observing period. Indeed, we find that the blackbody continues to be quite hot, plateauing around $kT_\mathrm{bb} \approx 0.2$~keV near peak X-ray luminosity and remaining close to this value during the X-ray decline.

The trends of the blackbody temperature with luminosity can be a useful gauge of the nature of the blackbody component. Blackbodies with constant emitting area and standard thin accretion disks are expected to follow $L \propto T^4$ \citep{Shakura1973}, whereas advection-dominated disks are expected to deviate from this trend and follow $L \propto T^2$ \citep[e.g.][]{Watarai2000}. Contrarily, the soft excess, both in typical AGN and the pre-outburst spectrum of 1ES~1927+654, is roughly constant in temperature, with $kT_\mathrm{bb} \approx 100-150$~eV and is independent of accretion rate and black hole mass \citep[e.g.][]{Gierlinski2004,Piconcelli2005,Miniutti2009}. 

We utilize these differences in expected behavior of the blackbody temperature to roughly assess the changes in the accretion state of 1ES~1927+654 over the course of the outburst. In the bottom panel of Figure \ref{fig:tempevol}, we show the luminosity of the blackbody as a function of its temperature, where the luminosity is determined using the normalization of the blackbody in XSPEC, given by $K_\mathrm{bb} = L_{39} / (D_{10} (1 + z))^2$, where $L_{39}$ is blackbody luminosity in units of $10^{39}$ erg s$^{-1}$ and $D_{10}$ is the distance to the source in units of 10 kpc. We color code the data by the phases in Figure \ref{fig:lumin} and fit the data in Phases 1 and 3 to the relationship $L \propto T^n$ using orthogonal distance regression (ODR). The results of fitting each individual group of data to $L \propto T^n$ are given in the legend of the bottom panel of Figure \ref{fig:tempevol}. We do not fit the data in Phase 2 as this corresponds to a relatively stable period where neither the luminosity nor the temperature of the blackbody component are changing enough to provide well-constrained fits to $L \propto T^n$. For the Phase 3 data, we excluded three data points at late time with low temperatures ($kT_\mathrm{bb} \approx 0.05$~keV) from the orange data points as these are likely to be from a different physical component of the system (the disk rather than the soft excess).

The data from Phase 1, shown in blue in Figure \ref{fig:tempevol}, are inconsistent with the standard $L \propto T^4$ picture for a thin accretion disk or constant emitting area blackbody. The slope is closer to the expected $L \propto T^2$ relationship for an advection dominated disk, although is still shallower. In combination with the hot blackbody temperature, this is indicative of a hot super-Eddington inner accretion flow. On the other hand, the late-time Phase 3 data are fit with an extremely steep slope in the $L-T$ plane, indicative of a temperature that is close to constant with luminosity. This could be the result of a rapidly shrinking effective blackbody radius or could be a phenomenological manifestation of the soft excess in AGN, given the relatively constant, high temperature relative to standard accretion disks. However, in this period there are still more significant deviations in the blackbody temperature than typically seen in AGN soft excesses, which could be the result of continued build up of the soft excess. Further monitoring of the source in its post-outburst state is necessary to determine the nature of the late-time blackbody component.

\subsection{Return of the Power Law Component and Softer-when-Brighter Behavior} \label{subsec:powerlaw}

\begin{figure}[t!]
    \centering
    \includegraphics[width=8.6cm]{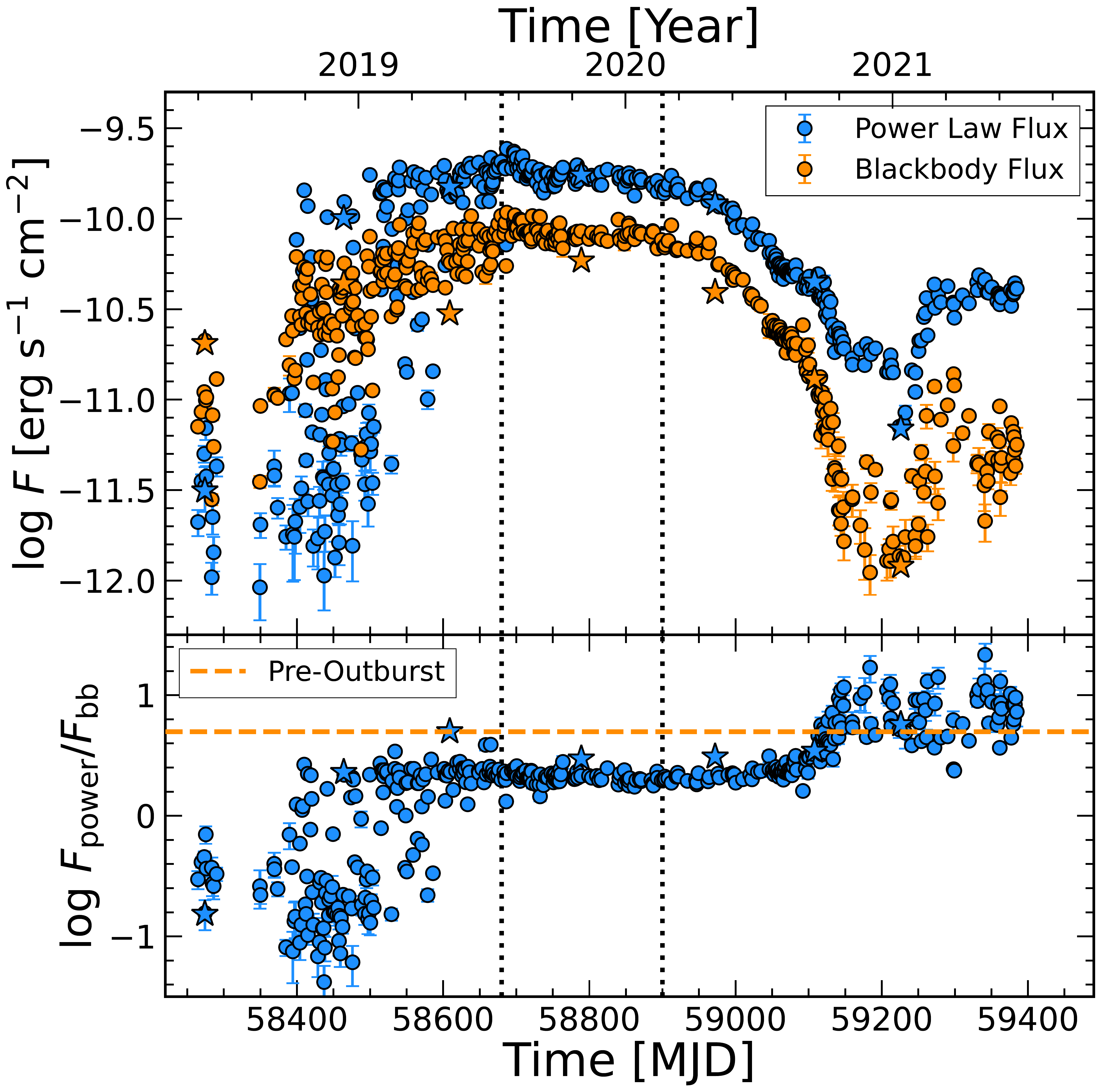}
    \caption{\textit{Top:} Power law (blue) and blackbody (orange) flux evolution in 1ES~1927+654. Both fluxes are computed in the 0.3-10 keV band. The stars show the \textit{XMM-Newton}/\textit{NuSTAR} data and the circles are the \textit{NICER} data. For visual clarity, we do not plot any data points with an uncertainty greater than 20\% on the either flux measurement. The black dotted lines show the phase boundaries defined in Figure \ref{fig:lumin}. \textit{Bottom:} Evolution of the ratio of power law to blackbody fluxes, with phase boundaries again overplotted. The stars show the \textit{XMM-Newton}/\textit{NuSTAR} data and the circles are the \textit{NICER} data. The pre-outburst flux ratio is shown with an orange dashed line.}
    \label{fig:fluxes}
\end{figure}

When 1ES~1927+654 was originally observed in X-rays following the optical transient, the power law component commonly seen in AGN was extremely weak and shortly disappeared as the luminosity dropped \citep{Ricci2020,Ricci2021}. Using our phenomenological model for 1ES~1927+654, we tracked the evolution of the power law component as it returned to the spectrum during the X-ray rise. We show the evolution of the flux of the power law component in the top panel of Figure \ref{fig:fluxes}, and in the bottom panel of Figure \ref{fig:lumin}, we show the evolution of the photon index of the power law. In Figure \ref{fig:fluxes}, we also show the blackbody flux in the top panel and the ratio of power law to blackbody flux in the bottom panel to assess the evolution of the dominant component of the spectrum. 

During the rise in X-ray luminosity in Phase 1, the power law component began to reappear and increase drastically in flux, exhibiting rapid variability similar to the observed variability in the X-ray light curve in the top panel of Figure \ref{fig:lumin}. Despite significant changes in flux, when we can constrain the photon index during Phase 1 and Phase 2, it remained relatively constant with $\Gamma \approx 3.5$, which is much steeper than both the pre-outburst spectrum ($\Gamma \approx 2.4$) and the typical photon index in AGN \citep[$\Gamma \approx 1.8$; e.g.][]{Ricci2017}. Then, when the X-ray luminosity begins to drops at the beginning of Phase 3, both the power law flux and photon index decrease steadily, ultimately returning to close to the pre-outburst flux and photon index from the 2011 \textit{XMM-Newton} spectrum. Given the negligible power law emission in X-ray spectra of the early Phase 1 observations, this rapid evolution back to the pre-outburst state suggests that the formation of the X-ray corona in AGN can be a rapid process. We further discuss the implications for these findings on the evolution of the X-ray corona in Section \ref{subsec:disc_corona}.

\begin{figure}[t!]
    \centering
    \includegraphics[width=8.6cm]{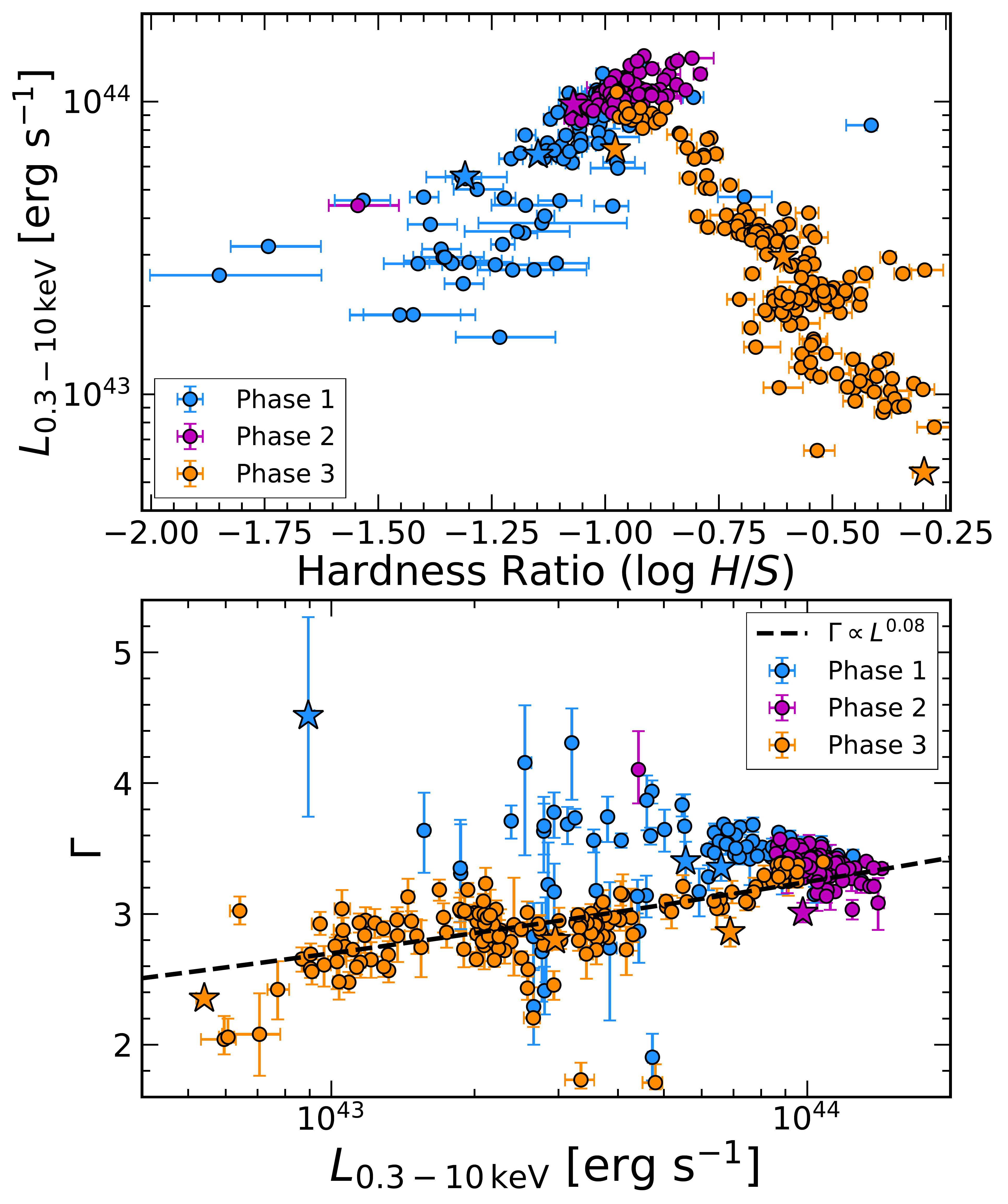}
    \caption{\textit{Top:} Hardness-intensity diagram for 1ES~1927+654 computed using model-based fluxes in 2-10~keV for the hard band and in 0.3-2~keV for the soft band. The stars show the \textit{XMM-Newton}/\textit{NuSTAR} data and the circles are the \textit{NICER} data. Data points are colored based on the phases identified in Figure \ref{fig:lumin}, with blue, magenta, and orange corresponding to Phase 1, 2, and 3 respectively. For visual clarity, we do not plot any data points with an uncertainty greater than 20\% on the hardness ratio. We also exclude the data points with $\Gamma$ frozen (as detailed in Section \ref{subsec:phenom_model}) as their 2-10~keV flux is poorly determined by the modeling. The softer-when-brighter-behavior associated with the corona is evident at late times in the Phase 3 data. \textit{Bottom:} Relationship between photon index and X-ray luminosity, colored again by time. The stars show the \textit{XMM-Newton}/\textit{NuSTAR} data and the circles are the \textit{NICER} data. We again exclude data points with $\Gamma$ frozen. The late-time data follow $\Gamma \propto L_X^{0.08}$, shown with the black dashed line, as found for standard AGN in \citet{Sobolewska2009}.}
    \label{fig:hid}
\end{figure}

Combining the decreasing photon index (i.e. the hardening of the X-ray spectrum) with the decreasing luminosity of 1ES~1927+654 during the Phase 3 observations leads to the standard ``softer-when-brighter" behavior that is typically exhibited by AGN \citep[e.g.][]{Shemmer2006,Sobolewska2009}. The top panel of Figure \ref{fig:hid} shows the luminosity versus hardness ratio, with the points colored by the phase from Figure \ref{fig:lumin}, highlighting the softer-when-brighter behavior at late times and the stark contrast to the early X-ray observations, which showed ``harder-when-brighter" during the X-ray rise \citep{Ricci2021}. Interestingly, harder-when-brighter behavior is also a commonly observed trend during the flares of quasi-periodic eruptions \citep[QPEs; e.g.][]{Miniutti2019,Giustini2020,Arcodia2021,Chakraborty2021} and in some ultra-luminous X-ray sources \citep[ULXs; e.g. NGC 247 ULX-1;][]{D'Ai2021}. We discuss this association further in Section \ref{subsec:disc_compare}, and compare the rapid variability during the X-ray rise to QPE behavior. 

To investigate the nature of the late-time softer-when-brighter behavior, we looked into the relationship between the photon index and luminosity, whose correlation in standard AGN indicate that this behavior is driven by the X-ray corona \cite[e.g][]{Sobolewska2009}. The bottom panel of Figure \ref{fig:hid} shows the photon index versus luminosity relationship, colored again by the phases identified in Figure \ref{fig:lumin}. In the late-time Phase 3 data, a clear trend is found where the photon index decreases for decreasing luminosity, consistent with what other standard AGN studies have found.

One possibility to explain the transition from harder-when-brighter to softer-when-brighter is that the dominant component of the spectrum has switched from thermal to Comptonized. The ratio of the power law to blackbody flux, shown in the bottom panel of Figure \ref{fig:fluxes}, is extremely low at early times, constant during much the peak X-ray luminosity phase and beginning of the X-ray decline, and increases later on in the X-ray decline. This implies that the early harder-when-brighter behavior is likely due to dominant blackbody emission and an increasing ratio of power law to blackbody flux. Once the source has reached its peak X-ray luminosity around an MJD of 58680, the ratio of the fluxes remains remarkably constant. Then at late times, both the decreasing photon index of the power law and the increase in the flux ratio of power law to blackbody flux likely contribute to the softer-when-brighter behavior.

\vspace{1cm}

\subsection{Similarity of Latest Observations to Pre-Outburst Observations} \label{subsec:comp2arch}

\begin{figure}[t!]
    \centering
    \includegraphics[width=8.6cm]{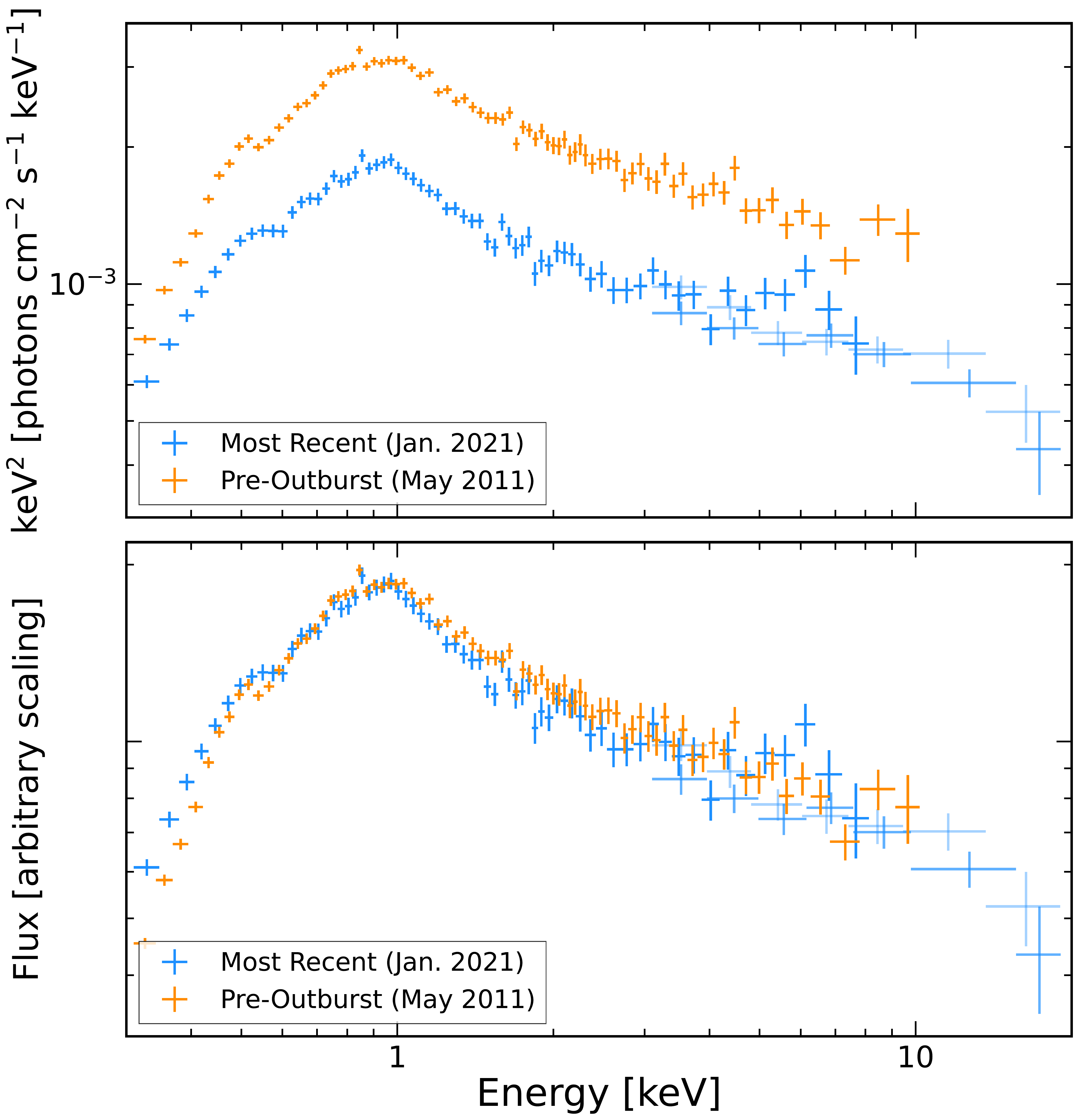}
    \caption{Comparison of the spectra from the pre-outburst 2011 \textit{XMM-Newton} observation (orange) and the most recent joint \textit{XMM-Newton}/\textit{NuSTAR} observation taken in January 2021 (blue). \textit{NuSTAR} FPMA and FPMB spectra from the January 2021 observation are in lighter shades of blue. Both spectra are unfolded using a $\Gamma = 0$ power law and have been rebinned for visual purposes. \textit{Top:} Unfolded spectra for both observations. \textit{Bottom:} Unfolded spectra have been renormalized such that they have the same 0.3-10 keV flux.}
    \label{fig:ep07_compare}
\end{figure}

Despite significant changes in the spectral shape and X-ray luminosity over the course of the past 3 years, the most recent observations of 1ES~1927+654 look remarkably similar to the pre-outburst \textit{XMM-Newton} observations from May 2011. In Figure \ref{fig:ep07_compare}, we show a comparison between the two spectra, highlighting the similarity of their spectral shapes. Pre-outburst observations spanning from 1990-2011 with \textit{XMM-Newton}, \textit{Suzaku}, \textit{Chandra}, and \textit{ROSAT} all showed a relatively steep X-ray spectrum ($\Gamma \approx 2.4-2.7$), with a rather hot soft excess and little obscuration \citep{Boller2003, Gallo2013}. The best phenomenological fit in \citet{Gallo2013} to the pre-outburst 2011 observation includes a $\Gamma = 2.39\pm0.04$ power law and $kT = 170\pm5$~eV blackbody. Both relativistically blurred reflection and ionized outflows, two physical models for the soft excess, provide a good fit to the pre-outburst spectrum \citep{Gallo2013}. The narrow Fe K$\alpha$ line, which is a nearly ubiquitous feature in AGN, was notably not detected in the 2011 spectrum. The upper limit on the equivalent width was significantly below what would be expected based on the X-ray Baldwin effect, whereby the Fe K$\alpha$ equivalent width is inversely correlated with the X-ray luminosity \citep[e.g.][]{Iwasawa1993,Page2004,Bianchi2007,Ricci2014}, suggesting that the circumnuclear environment was devoid of gas and dust.

We find that the most recent simultaneous \textit{XMM-Newton}/\textit{NuSTAR} observation of 1ES~1927+654 can also be fit well by a power law and blackbody model with very similar fit parameters, finding a photon index of $\Gamma = 2.35 \pm 0.03$ and a blackbody temperature of $kT_\mathrm{bb} = 183_{-6}^{+7}$~eV. This gives a satisfactory fit with $\chi^2_\nu/\nu = 1.09/655$, but this fit can be improved by including an additional low temperature blackbody that could be from the accretion disk with $kT_\mathrm{bb} = 40 \pm 4$~eV, giving $\chi^2_\nu/\nu = 1.00/653$ ($\Delta\chi^2 = 63$ for 2 additional degrees of freedom). We find that a number of physical models for the soft excess can provide a good fit to this spectrum, indicating that the hotter blackbody component is likely associated with the soft excess as in the pre-outburst spectrum. Details of the physical modeling of the soft excess are outlined further in Section \ref{subsec:jan2021}. 

Given the softness of the spectrum, the source is only dominant over the \textit{NuSTAR} background up to 20~keV, making constraining the cutoff energy of the corona difficult. The above phenomenological modeling suggests a cutoff energy of $E_\mathrm{cut} \gtrsim 15$~keV, although this value is somewhat dependent on the choice of model. We further explore the cutoff energy evolution and what this tells us about the formation and heating processes and timescales in the corona in Section \ref{subsec:disc_corona}, but note that the latest \textit{XMM-Newton}/\textit{NuSTAR} observations reveal a cutoff energy that is not unlike other AGN corona, suggesting that the source returned to close to its pre-outburst AGN state. We also find that the \textit{XMM-Newton} residuals show a hint of a feature around 6.4~keV in this observation, but no such feature seems to exist in the \textit{NuSTAR} data with lower energy resolution. Adding an additional Gaussian line at 6.4~keV to an \textit{XMM-Newton} only fit is significant at the 97\% confidence level (using an F-test to provide an approximation of the significance), but when including the \textit{NuSTAR} data we find that an additional Gaussian component at 6.4~keV does not provide any statistically-significant improvement to the fit. 

Following the latest \textit{XMM-Newton}/\textit{NuSTAR} observations, the \textit{NICER} data reveal that 1ES~1927+654 brightened again by a factor of approximately four in 0.3-10~keV luminosity over the span of approximately two months. Archival flux values from \textit{ROSAT} and \textit{Chandra} observations in the 1990s and early 2000s reveal that this flux increase is not outside of the realm of normal for 1ES~1927+654 \citep[see Figure 9 from][]{Gallo2013}. This second rise is notably different than the initial X-ray rise in early 2018 as it is not rapidly variable, instead following the smooth nature seen in the X-ray decline, which could suggest that this change is more indicative of standard stochastic AGN variability. The latest \textit{NICER} data studied in this work are still fit well by the simple power law plus blackbody model discussed in Section \ref{subsec:phenom_model} with an average fit statistic of $\chi^2_\nu = 1.13$. Compared to the latest \textit{XMM-Newton}/\textit{NuSTAR} fits, the latest \textit{NICER} observations reveal that the average power law is steeper with $\Gamma \approx 2.9$, consistent with the typical softer-when-brighter behavior exhibited by AGN, and the average temperature of the blackbody is lower with $kT_\mathrm{bb} \approx 150$~eV. 



\section{Physically-Motivated Modeling with Blurred Reflection} \label{sec:physmod}

One of the most striking features in the early spectra of 1ES~1927+654 was a broad emission-like excess around 1~keV. In Section \ref{subsec:disapp_1keV} we modeled the 1~keV feature as a simple Gaussian emission line and showed that the equivalent width drops when the source reaches its peak X-ray luminosity (near the Phase 1--Phase 2 boundary in Figure \ref{fig:lumin}). In this section, we explore a potential physical model for the 1~keV feature, namely reflection from a single-temperature blackbody irradiating spectrum, that can also account for broad residuals in the peak X-ray luminosity spectrum. We focus on three \textit{XMM-Newton}/\textit{NuSTAR} epochs which correspond to each of the three unique evolutionary phases highlighted in Figure \ref{fig:lumin}. Specifically, we use the data from \textit{XMM-Newton}/\textit{NuSTAR} Epochs 1, 4, and 7 as these are representative extremes in each of the regimes.

\subsection{June 2018 XMM-Newton Observation and the 1~keV Feature} \label{subsec:june2018}

The first \textit{XMM-Newton} observation of 1ES~1927+654 after the optical outburst began (June 2018, Epoch 1) shows the strongest 1~keV feature, as can be seen from the ratio plot in Figure \ref{fig:1keV_eqw}. The residuals resemble those of the ultrafast outflow in ASASSN-14li, although at a higher energy \citep{Kara2018}. To test this idea, we performed photoionized absorption modeling of the feature using an \textsc{xstar} absorption grid, which assumed that the ionizing spectrum was a blackbody with $kT_\mathrm{bb} = 100$ eV. However, we were unable to fit the data with a single absorber, finding a poor fit with $\chi^2_\nu > 2$. A model with two separate absorption components provides a slightly better fit ($\chi^2_\nu \approx 1.5$), but requires extreme parameters, including a blueshift of $z \approx -0.5$ for one of the absorbers, which would greatly exceed any observed outflow velocity seen in other accreting sources. An alternative explanation is that the feature is seen instead in emission. The width of the feature is on the order of 0.1~keV \citep[and even appears broad in the \textit{XMM-Newton} RGS data as seen in][]{Ricci2021}, suggesting emission from relatively close to the black hole and motivating the use of blurred reflection to model the 1~keV feature. 

Reflection modeling in AGN usually assumes that the hot, optically thin corona irradiates the disk with a cutoff power law or thermally Comptonized irradiating spectrum \citep[e.g. \texttt{xillver}, \texttt{xillverCp};][]{Garcia2010,Garcia2013}. This is not applicable in the early X-ray observations of 1ES~1927+654 given the extremely weak power law component and dominant soft thermal component. To model reflection we therefore used a new model, \texttt{xillverTDE}, which models reflection from a single-temperature blackbody irradiating spectrum, reminiscent of the thermal continuum in the early X-ray spectra of 1ES~1927+654. The \texttt{xillverTDE} model is a new flavor of the \texttt{xillver} suite of models, which utilizes the largest atomic database and most accurate radiative transfer calculations for reflection. The model has seven free parameters, including the incident blackbody temperature, iron abundance, ionization parameter, disk density, inclination, redshift, and normalization.

Blackbody illuminating spectra are not only useful for modeling reflection in the soft thermal spectra in TDEs, but are also a key ingredient in modeling reflection from the surface or boundary layer of neutron stars and returning radiation in black hole accretion systems. Recently, a flavor of the \texttt{xillver} models called \texttt{xillverNS} has been developed to model the reflected emission from a single-temperature blackbody illuminating spectrum in neutron stars \citep{Garcia2022}. Developed for X-ray binary systems, this model includes a hotter temperature blackbody than in \texttt{xillverTDE}, which includes cooler temperatures in the range $kT_\mathrm{bb} = 0.03-0.3$~keV. \texttt{xillverNS} has proven extremely valuable for probing neutron star properties through reflection modeling \citep[e.g.][]{Ludlam2018,Ludlam2019,Ludlam2020} and as a probe for returning radiation in black hole X-ray binaries \citep[e.g.][]{Connors2020,Connors2021}. 

\begin{figure*}
    \centering
    \includegraphics[width=\textwidth]{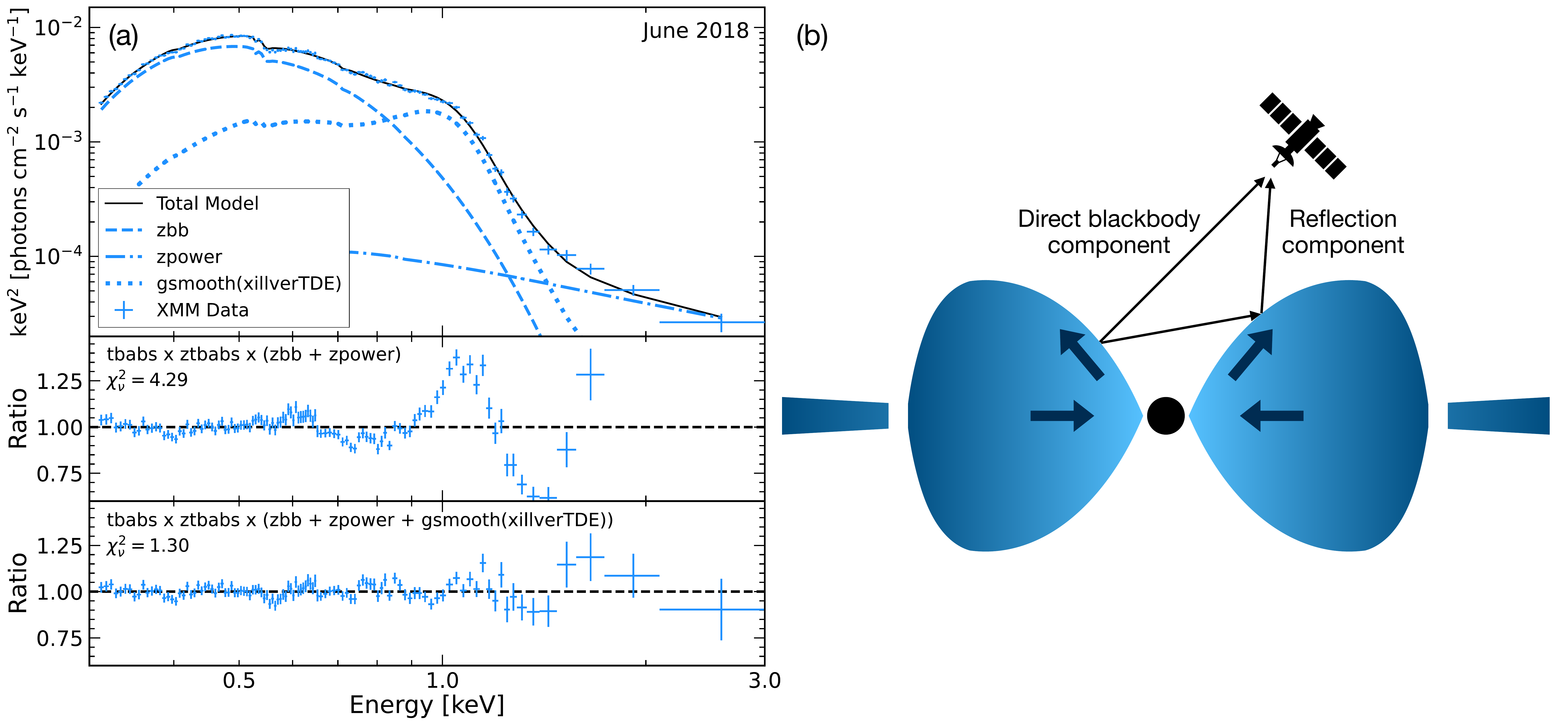}
    \caption{\textit{Left:} First \textit{XMM-Newton} observation fit to the model \texttt{tbabs*ztbabs*(zbb+zpower+gsmooth(xillverTDE))}. The top panel shows the unfolded spectrum with the three additive spectral components. The middle panel shows the ratio plot to a simple blackbody and power law fit without the \texttt{xillverTDE} component to highlight the vast improvement that the component has on the model. The bottom panel shows the ratio of the data to the model including \texttt{xillverTDE}. Data have been rebinned for visual purposes. \textit{Right:} Schematic showing the \texttt{xillverTDE} reflection modeling off of an optically-thick, super-Eddington outflow. }
    \label{fig:xillverTDE}
\end{figure*}

To test whether reflection can provide a good description of the 1~keV emission in 1ES~1927+654, we utilize the \texttt{xillverTDE} model to fit the Epoch 1 \textit{XMM-Newton} data. In XSPEC notation, the model used is \texttt{tbabs*ztbabs*(zbb+zpower+gsmooth(xillverTDE))}. Recent simulations suggest that line profiles from super-Eddington disks are more symmetric and blueshifted compared to those from standard thin accretion disks \citep[e.g.][]{Thomsen2019,Thomsen2022}. Although the X-ray luminosity is about 10\% the Eddington limit for a $10^6\,M_\odot$ black hole, multi-wavelength SED modeling suggests that the overall luminosity is super-Eddington (Li et al. in prep).  Hence, we allowed the \texttt{xillverTDE} component to be blueshifted and smooth the model with a constant velocity broadening to avoid the assumption of a standard thin accretion disk that is invoked in relativistic convolution models like \texttt{relconv} \citep{Dauser2010}. We also link the incident blackbody temperature with the blackbody of the \texttt{xillverTDE} model. The disk may see a slightly modified blackbody due to strong gravity effects around the black hole, but freeing the blackbody temperature does not provide a significantly improved fit. Likewise, most of the reflected emission in the 1~keV line is likely coming from close to the black hole, and thus a single-temperature approximation for a multi-temperature disk blackbody should be reasonable. We freeze the iron abundance in \texttt{xillverTDE} to solar and the inclination at $i = 45^\circ$ as the fit is not sensitive to inclination without relativistic blurring included. Leaving the inclination free does not significantly improve the fit, although the inclination angle is constrained to be $\lesssim 50^\circ$. We fit for the ionization parameter, disk density, redshift, and normalization and report the result of our fits in Table \ref{tab:xillverTDE}.

In the left panel of Figure \ref{fig:xillverTDE}, we show the results of using \texttt{xillverTDE} to model the 1~keV excess in the first \textit{XMM-Newton} observation. We find that a blueshifted reflection component with $z = -0.33$ provides a good fit to the data, giving $\chi^2_\nu/\nu = 1.30/216$ and significantly improving the residuals around 1~keV. We note that an equally good fit can be obtained using the \texttt{relconv} convolution model for relativistic blurring instead of Gaussian smoothing, but the fit requires a high iron abundance and a high inclination to achieve a significant blueshift on the \texttt{xillverTDE} model. In addition, the \texttt{relconv} fit requires an inner disk radius very close to the innermost stable circular orbit (ISCO), which is potentially inconsistent with an edge-on, geometrically thick super-Eddington inner accretion flow. In Section \ref{subsec:disc_phys_xillverTDE}, we discuss further the implications of this modeling and suggest that this could be reprocessed emission off of a geometrically thick accretion disk from a super-Eddington accretion flow. Another similar scenario has been invoked for the emission lines in the high-resolution \textit{XMM-Newton} RGS spectrum of the rapidly accreting AGN 1H 1934-063 by \citet{Xu2022}.

\subsection{November 2019 XMM-Newton/NuSTAR Observation} \label{subsec:nov2019}

\begin{figure*}[t!]
    \centering
    \includegraphics[width=\textwidth]{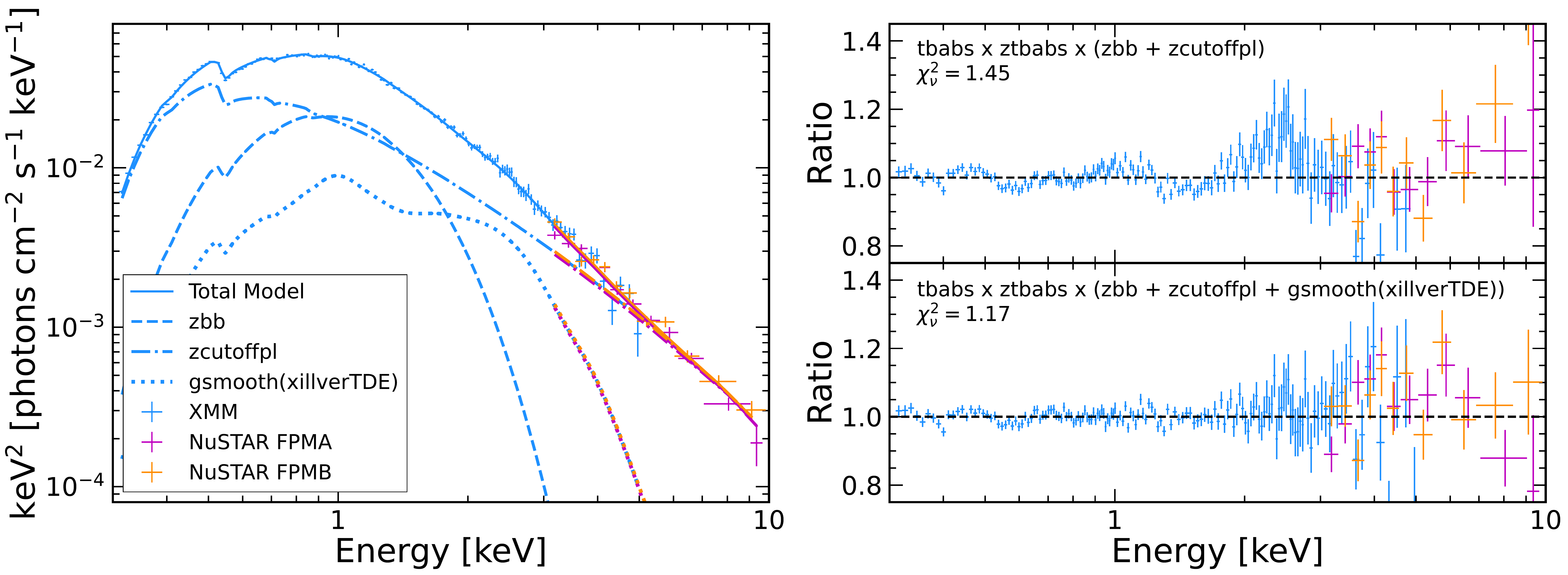}
    \caption{\textit{Left:} Unfolded spectrum from the November 2019 simultaneous \textit{XMM-Newton}/\textit{NuSTAR} observation of 1ES~1927+654 fit to the velocity-broadened \texttt{xillverTDE} model. Data have been rebinned for visual purposes. \textit{Top Right:} Ratio plot to the simple cutoff power law and blackbody model used in phenomenological modeling, highlighting the need for additional components in the spectral model. \textit{Bottom Right:} Ratio plot for the reflection model, including the \texttt{xillverTDE} model from a blackbody irradiating spectrum. }
    \label{fig:ep4}
\end{figure*}

As shown in Figure \ref{fig:1keV_eqw}, the equivalent width of the 1~keV line significantly decreases when the source reaches its peak luminosity and does not strengthen when the luminosity starts to decrease. This can be attributed both to the increased continuum flux and also a slight decrease in the line flux. Despite this drop in equivalent width and line flux measured in the phenomenological modeling, in the November 2019 \textit{XMM-Newton} spectrum (Epoch 4, taken when the source was at its peak X-ray luminosity), there are some broad residuals around 1~keV and 2.5~keV in the simple continuum fit. We show the ratio plot for this spectrum relative to a simple cutoff power law and blackbody model in the top right panel of Figure \ref{fig:ep4}, which has notable residuals and provides a relatively poor fit to this high quality spectrum ($\chi^2_\nu / \nu = 1.45/736$). 

The source is almost certainly super-Eddington during this period, given the high X-ray luminosity (see Figure \ref{fig:lumin}), and therefore likely still has a geometrically thick inner accretion flow. Thus, the blackbody-based reflection component from the Epoch 1 \textit{XMM-Newton} observation could still be present in the Epoch 4 spectrum. We test this by using the same \texttt{xillverTDE} model, with the blackbody temperatures linked, solar abundances, a free blueshift parameter, and a constant velocity broadening, as with the Epoch 1 spectrum. This model is able to reproduce the two broad residual components at roughly 1~keV and 2.5~keV and provides a significant improvement to the spectral fit with $\chi^2 / \nu = 1.17/731$. In Figure \ref{fig:ep4}, we show the resulting ratio plot in the bottom right panel and the unfolded spectrum with the model components in the left panel. We also report the resulting fit parameters in Table \ref{tab:xillverTDE} and note that the density, blueshift, and broadening are comparable with the fit parameters from the Epoch 1 spectrum. The ionization parameter is larger than the Epoch 1 value, which is likely indicative of a higher ionizing luminosity and suggests that the change in the 1~keV feature seen in Figure \ref{fig:1keV_eqw} and discussed in Section \ref{subsec:disapp_1keV} may be an ionization effect. We note that as with Epoch 1, a similarly good fit to Epoch 4 spectrum can be obtained by blurring the \texttt{xillverTDE} model with \texttt{relconv}, with the same high inclination and high iron abundance required as in Epoch 1.

However, the Epoch 4 X-ray spectrum is now dominated by a relatively steep power law component, so there may be additional coronal reflection components that are not included in this modeling. Therefore, in addition to a blackbody-irradiating reflection model, we also tried including Gaussian-smoothed \texttt{xillverD}, a standard reflection model with variable density from a power law ionizing spectrum \citep{Garcia2016}. We find that this model can also provide a good fit to the spectrum ($\chi^2 / \nu = 1.15/729$), but that the addition of coronal reflection provides only marginal improvement in the spectral fit ($\Delta \chi^2 \approx 10$ for 2 fewer dof). Thus, we cannot rule out the possibility that the corona also produces reflection features, but our modeling does not statistically require the additional model component.

\subsection{January 2021 XMM-Newton/NuSTAR Observation} \label{subsec:jan2021}

In Section \ref{subsec:comp2arch}, we showed with phenomenological modeling that 1ES~1927+654 had returned to a state very similar to its pre-outburst state from the 2011 \textit{XMM-Newton} observation \citet{Gallo2013}, with a strong soft excess and a power law component with $\Gamma \approx 2.4$. Given the success of fitting the Epoch 1 and 4 data with reflection models, here we explore whether relativistic reflection can be sufficient to model the most recent simultaneous \textit{XMM-Newton}/\textit{NuSTAR} observation (January 2021, Epoch 7). Despite the lack of a strong Fe K feature in the X-ray spectrum of 1ES~1927+654, relativistically blurred reflection is a popular physically-motivated model for the soft excess \citep[e.g.][]{Crummy2006,Fabian2009,Garcia2019}. Moreover, the similarity of the X-ray spectrum to its pre-outburst state with a strong power law component motivates the use of a standard reflection model from a power law ionizing spectrum and relativistic blurring from a thin accretion disk. 

\begin{figure}[t!]
    \centering
    \includegraphics[width=8.6cm]{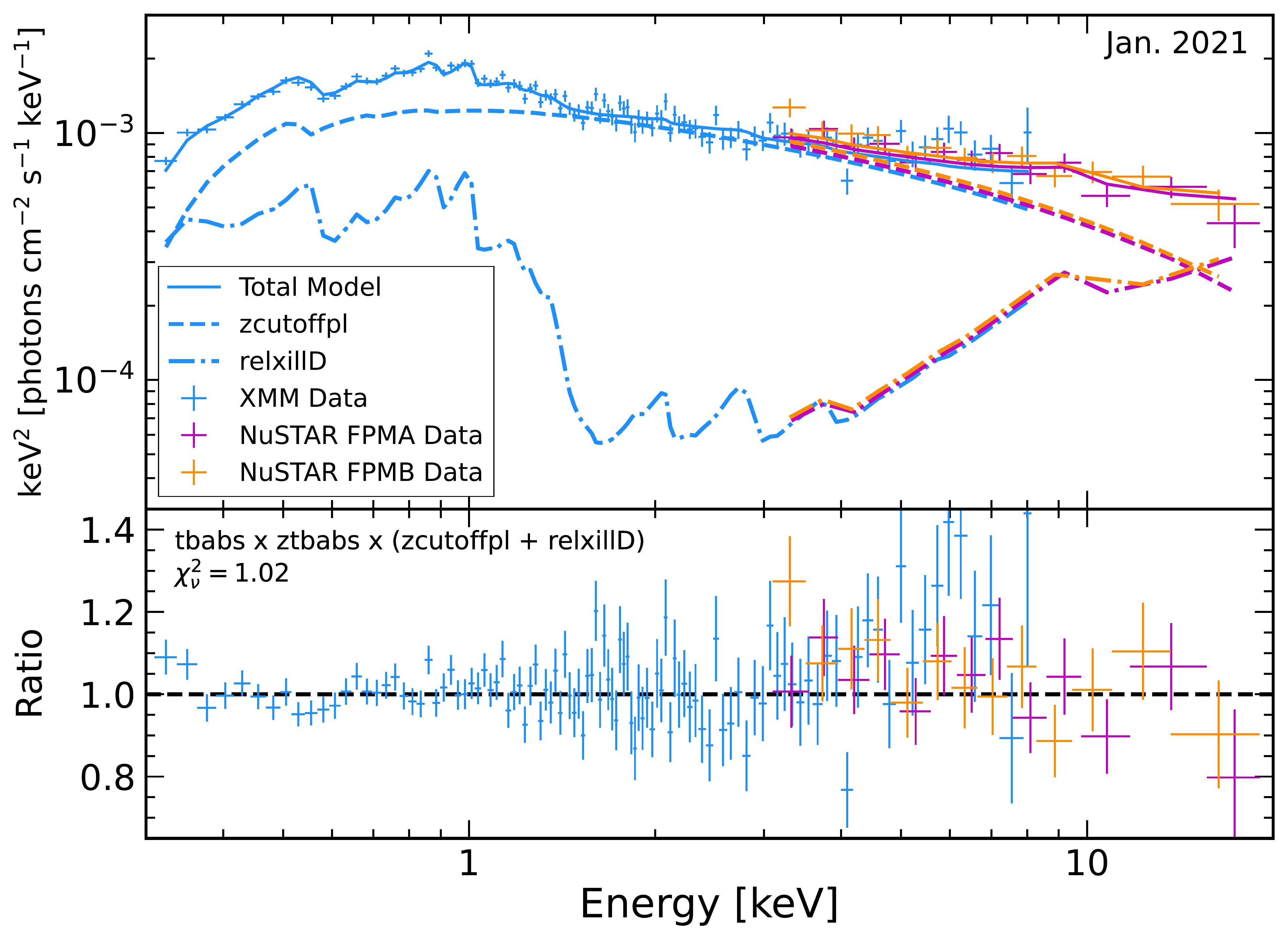}
    \caption{\textit{Top:} Unfolded spectrum from the January 2021 simultaneous \textit{XMM-Newton}/\textit{NuSTAR} observation of 1ES~1927+654 fit to the cutoff power law and standard relativistic reflection model. \textit{Bottom:} Ratio plot for the same model. Data have been rebinned for visual purposes.}
    \label{fig:ep7}
\end{figure}

We model reflection using the high density model \texttt{relxillD}, a flavor of the \texttt{relxill} suite of models which includes relativistic blurring from a thin accretion disk \citep{Dauser2014,Garcia2014}. High density reflection models have been shown to produce a stronger soft excess due to free-free processes dominating heating and leading to an increase in the temperature of the top layer of the disk \citep{Garcia2016}. As with the previous reflection modeling, we leave the ionization parameter, disk density, and normalization as free fit parameters. We use relativistic blurring from a thin accretion disk instead of a constant velocity broadening, so we also fit for the inclination, emissivity index, and spin of the black hole. We fix the inner edge of the accretion disk to the ISCO, which scales with the spin of the black hole, and fix the outer radius to $R_\mathrm{out} = 400 R_g$. We find that the relativistic reflection modeling can provide a good fit to the soft excess with $\chi^2_\nu / \nu = 1.02/650$. We show the results of this modeling in Figure \ref{fig:ep7}, and report the fit values for the reflection model in Table \ref{tab:xillverTDE}. 

The current \texttt{relxillD} models have a fixed cutoff energy of $E_\mathrm{cut} = 300$~keV, while our fits suggest a lower cutoff energy. We thus found a degeneracy between the strength of the reflection component near the Compton hump and the cutoff energy of the power law component. In Figure \ref{fig:ep7} and Table \ref{tab:xillverTDE}, we show the high density, low cutoff energy fit but note that this is degenerate with a lower density, higher cutoff energy fit ($n \approx 10^{16}$ cm$^{-3}$, $E_\mathrm{cut} \approx 50$ keV). To test this degeneracy, we also fit the data with \texttt{relxillDCp}, which uses a thermal Comptonization illuminating spectrum with a variable coronal temperature. With \texttt{relxillDCp} and a corresponding thermal Comptonization spectrum for the continuum (\texttt{nthcomp} model in XSPEC), we find a coronal temperature of $kT_e = 6.1_{-1.2}^{+1.6}$ keV and a relatively high density around $\log (n / \mathrm{cm}^{-3}) \approx 18$, which is consistent with the low cutoff energy that we report in Table \ref{tab:xillverTDE} for the \texttt{relxillD} modeling (with $E_\mathrm{cut} \approx 2-3\, kT_e$). Further hard X-ray observations will better constrain the coronal temperature and the nature of the potential Fe K line.

The model suggests an edge on geometry with maximal inclination angle, which is somewhat in tension with emission off of a funnel geometry interpretation for the 1~keV line in the early X-ray spectra. However, inclination mismatches are not uncommon in X-ray spectral fitting, especially for rapidly accreting black holes under the assumption of thin accretion disks \citep[see, for example, Section 4.2 of][and references therein]{Mundo2020}, and the inner flow may be misaligned from the outer disk in the early X-ray observations (see Section \ref{subsec:disc_phys_xillverTDE}). Weak iron line features at $\sim6-7$~keV are not uncommon in sources with a strong soft excess \citep[e.g.][]{Mallick2018,Garcia2019,Xu2021} and are consistent with high density reflection models, which predict that the high density gas becomes hotter due to enhanced free-free heating, thus increasing the broadening of the line \citep{Garcia2016}. Likewise, the lack of a strong iron line feature in 1ES~1927+654 could also be explained by an extended corona that is in turn further blurring reflection from the inner accretion disk via a second Comptonization \citep[e.g.][]{Steiner2017}. This would only be relevant in a relatively low inclination system where the line of sight to the inner accretion flow intersects the corona, as suggested by our modeling of the broad 1~keV feature. 

We also note that this relativistic reflection modeling is not the only possible model for the soft excess, which is a current mystery in the AGN X-ray community. Another common model for the soft excess is a warm, optically thick corona \citep[e.g.][]{Mehdipour2011,Done2012}. Often these two models cannot be distinguished from one another by statistics alone \citep[e.g.][]{Garcia2019,Ghosh2021,Xu2021}, adding to the confusion behind this feature. We find that the soft excess can also be modeled with a warm corona model, using \texttt{nthcomp} with a low electron temperature to model this component, which may alleviate the inclination mismatch as the late-time high inclination constraints are strongly dependent on the reflection modeling of the soft excess. Likewise, we also test an ionized outflow model, which was found to be consistent with the pre-outburst soft X-ray spectrum of 1ES~1927+654 if two ionized absorbers were included \citep{Gallo2013}. We find that a single ionized absorber (modeled with \texttt{zxipcf}) with $N_H \approx 10^{24}$ cm$^{-2}$, $\log\xi \approx 2.8$, and $z \approx -0.3$ can improve the soft X-ray fit in the Epoch 7 observation of 1ES~1927+654 with $\chi^2_\nu/\nu = 1.09/653$. An additional absorber does not significantly improve the fit, but as with the phenomenological modeling presented in Section \ref{subsec:comp2arch}, we find significant improvement with an additional low temperature blackbody component ($\Delta\chi^2 = 48$ for 2 additional dof, $kT_\mathrm{bb} \approx 30$ eV). All three models for the soft excess produce similar fit statistics, as is commonly found in other sources. Distinguishing between soft excess models is beyond the scope of this work.

\begin{deluxetable*}{c c c c c}

	\caption{Fit Results for the Blurred Reflection Modeling with \texttt{xillverTDE}} \label{tab:xillverTDE}
	
    \tablehead{\colhead{Model Component (XSPEC Model)} & \colhead{Parameter (Units)} & \colhead{Epoch 1\tablenotemark{a}} & \colhead{Epoch 4\tablenotemark{b}} & \colhead{Epoch 7\tablenotemark{c}}}

	\startdata
	Galactic absorption (\texttt{tbabs}) & $N_{H,\,\mathrm{gal}}$ ($10^{20}$ cm$^{-2}$) & 6.42\tablenotemark{f} & 6.42\tablenotemark{f} & 6.42\tablenotemark{f} \\
	Intrinsic absorption (\texttt{ztbabs}) & $N_H$ ($10^{20}$ cm$^{-2}$) & $2.6_{-0.5}^{+0.6}$ & $7.7_{-0.7}^{+0.7}$ & $1.8_{-1.7}^{+1.4}$ \\
	Blackbody (\texttt{zbb}) & $kT_\mathrm{bb}$ (eV) & $88_{-2}^{+2}$ & $205_{-2}^{+8}$  & -- \\
	& $F_{0.3-10\,\mathrm{keV,\, bb}}$ ($10^{-11}$ erg cm$^{-2}$ s$^{-1}$) & $1.8_{-0.1}^{+0.1}$ & $5.8_{-0.9}^{+0.2}$ & -- \\
	Cutoff Power Law (\texttt{zcutoffpl}) & $\Gamma$ & $3.4_{-0.7}^{+0.8}$ & $3.59_{-0.11}^{+0.12}$  & $2.30_{-0.10}^{+0.11}$ \\
	& $E_\mathrm{cut}$ (keV) & --\tablenotemark{d} & $8.1_{-2.1}^{+4.5}$ & $16.1_{-6.7}^{+28.7}$ \\
	& $F_{0.3-10\,\mathrm{keV,\, pl}}$ ($10^{-11}$ erg cm$^{-2}$ s$^{-1}$) & $0.06_{-0.03}^{+0.08}$ & $17.4_{-1.5}^{+1.2}$ & $0.67_{-0.04}^{+0.04}$ \\
	Blackbody Reflection (\texttt{xillverTDE}) & $\log(\xi$/erg cm s$^{-1}$) & $2.48_{-0.14}^{+0.01}$ & $3.00_{-0.57}^{+0.03}$ & -- \\
	& $\log (n$/cm$^{-3}$) & $18.9_{-0.6}^{+0.1}$ & $19.0_{-0.7}$\tablenotemark{e} & -- \\
	& $i$ (deg) & $45$\tablenotemark{f} & $45$\tablenotemark{f} & -- \\
	& $z$\tablenotemark{g} & $-0.33_{-0.01}^{+0.01}$ & $-0.33_{-0.01}^{+0.04}$ & -- \\
	& $F_{0.3-10\,\mathrm{keV,\, xillverTDE}}$ ($10^{-11}$ erg cm$^{-2}$ s$^{-1}$) & $0.45_{-0.03}^{+0.06}$ & $2.5_{-0.4}^{+0.6}$ & \\
	Corona Reflection (\texttt{relxillD}) & $\log(\xi$/erg cm s$^{-1}$) & -- & -- & $1.24_{-0.09}^{+0.20}$ \\
	& $\log (n$/cm$^{-3}$) & -- & -- & $18.3_{-0.4}^{+0.5}$ \\
	& $i$ (deg) & -- & -- & $87_{-3}\tablenotemark{e}$ \\
	& $q$ & -- & -- & $6.9_{-1.9}^{+3.1}$ \\
	& $a$ & -- & -- & $0.93_{-0.10}^{+0.068}$ \\
	& $F_{0.3-10\,\mathrm{keV,\, relxillD}}$ ($10^{-11}$ erg cm$^{-2}$ s$^{-1}$) & -- & -- & $0.25_{-0.07}^{+0.08}$ \\
	Velocity Broadening (\texttt{gsmooth}) & $v$ $(c)$ & $0.12_{-0.01}^{+0.01}$ & $0.10_{-0.03}^{+0.09}$ & -- \\
	Cross-Calibration (\texttt{const}) & $C_\mathrm{FPMA}$ & -- & $0.97_{-0.05}^{+0.05}$ & $1.04_{-0.06}^{+0.07}$ \\
	& $C_\mathrm{FPMB}$ & -- & $1.01_{-0.05}^{+0.05}$ & $1.08_{-0.07}^{+0.07}$ \\
	Fit Statistic & $\chi^2_\nu/\nu$ & 1.30/216 & 1.17/731 & 1.02/650 \\
    \enddata
    
    \tablenotetext{a}{June 2018, Model: \texttt{tbabs*ztbabs*(zbb+zpower+gsmooth(xillverTDE))}}
    \tablenotetext{b}{November 2019, Model: \texttt{tbabs*ztbabs*(zbb+zcutoffpl+gsmooth(xillverTDE))}}
    \tablenotetext{c}{January 2021, Model: \texttt{tbabs*ztbabs*(zcutoffpl+relxillD)}}
    \tablenotetext{d}{No cutoff was included in this fit, given that the data only extend to 3~keV.}
    \tablenotetext{e}{Parameter pegged at maximum value of the model.}
    \tablenotetext{f}{Parameter was fixed when fitting.}
    \tablenotetext{g}{Intrinsic blueshift, corrected for the cosmological redshift of the host galaxy.}
    
\end{deluxetable*}


\section{Discussion} \label{sec:discussion}

\subsection{A Relativistic Outflow Origin for the Broad 1~keV Line} \label{subsec:disc_phys_xillverTDE}

In Section \ref{sec:physmod}, we show that \texttt{xillverTDE}, a new flavor of the \texttt{xillver} reflection models with a single-temperature blackbody irradiating spectrum, can reproduce the broad 1~keV feature that is prominent in the early X-ray spectra of 1ES~1927+654. In this model, the 1~keV feature is primarily the result of velocity-broadened and blueshifted oxygen K-shell emission. The significant velocity broadening and blueshift required in our \texttt{xillverTDE} modeling suggests outflowing emission from close to the black hole. We show a schematic of how this emission could arise from the base of an outflow in a super-Eddington, geometrically thick inner accretion flow in the right panel of Figure \ref{fig:xillverTDE}. Simulations of super-Eddington accretion disks have shown that significant radiation pressure in the geometrically thick accretion disks causes optically thick winds to be launched from the disk \citep[e.g.][]{Ohsuga2009,Jiang2014,McKinney2014}. Likewise, some ULXs and TDEs, both of which are thought to be radiating close to or above the Eddington limit, have been shown to launch fast outflows \citep[e.g.][]{Middleton2014,Pinto2016,Walton2016,Kara2018,Kosec2018a,Kosec2018b,Pinto2021}. Thus, a geometrically thick accretion disk launching significant outflows is entirely plausible in 1ES~1927+654 and is supported by the modeling with \texttt{xillverTDE}.

Additionally, \citet{Ricci2020} first suggested that the outburst in 1ES~1927+654 was the result of a TDE in a pre-existing accretion disk. In this picture, the stellar debris hits the disk and produces shocks that cause the gas in the inner accretion flow to lose angular momentum and fall onto the black hole. This depletes the inner accretion flow and can thus cut off the energy supply to the corona. Hydrodynamic simulations of TDEs in AGN suggest that this depletion of the inner accretion flow happens in a super-Eddington manner, with a thick inner accretion disk formed \citep{Chan2019}. This optically and geometrically thick inner accretion disk could then easily irradiate itself, producing the reflected emission from a blackbody ionizing spectrum that we see in the early X-ray spectra as a broad 1~keV excess. In order to see this emission, however, we must have a sight line to the inner accretion flow. We suspect that to see the reflected emission off of this geometrically thick outflow we would require a relatively face-on accretion geometry, although the exact constraints on the inclination angle depend on the scale height $H/R$ of the disk, which is dependent on the Eddington ratio and hard to determine precisely. 

Another possible way to get a sight line to the inner accretion flow is with a warped inner accretion disk. In TDEs, the misalignment between the angular momentum of the tidally disrupted star and the black hole in combination with Lens-Thirring precession has been shown theoretically to lead to warped accretion disks \citep[e.g.][]{Stone2012,Franchini2016}. These warped accretion disk models have been suggested to explain the quasi-periodic nature of the early X-ray light curve of the jetted TDE, Swift J1644 \citep[e.g.][]{Reis2012,Lei2013}. Similar physics could be at play in 1ES~1927+654 with a TDE that is misaligned with an existing accretion disk, leading to a warped inner accretion flow relative to the outer accretion disk. This misalignment would persist until angular momentum transport aligned the disk and the black hole spin axis. Thus, this could allow the outer accretion disk to have a higher inclination angle as suggested by reflection modeling of the pre-outburst and latest observations, while still allowing us to see into the inner accretion flow to see the broad 1~keV line during the early observations. The alignment timescale depends on the properties of the system, but for a $10^6 \, M_\odot$ black hole, solid body precession of the inner accretion flow could persist for on the order of one year \citep[e.g.][]{Stone2012,Franchini2016}, similar to the time in which 1ES~1927+654 exhibits rapid variability.

The success of \texttt{xillverTDE} goes beyond just the first \textit{XMM-Newton} observation. As shown in Figure \ref{fig:ep4}, the peak X-ray luminosity spectral fit is also greatly improved by including reflection from the blackbody. This could suggest that blackbody reflection is a key component of many super-Eddington accretors and super-soft X-ray sources where the thermal component dominates the X-ray continuum. Likewise, a similar scenario of soft reflection at the base of a wind has been invoked to explain emission features in the narrow line Seyfert 1 1H 1934-063, which is accreting near or above the Eddington limit \citep{Xu2022}. In addition to rapidly accreting AGN, TDEs in low mass supermassive black holes ($M \lesssim 10^7\, M_\odot$) are known to have super-Eddington fallback rates and very soft, thermal X-ray spectra, making them a perfect probe of this model. Broad soft X-ray features similar to the 1~keV line in 1ES~1927+654 are indeed present in other TDEs \citep[e.g.][]{Kara2018}, and future work to test this model on other X-ray detected TDEs with broad lines in their X-ray spectra is currently in progress (Masterson et al. in prep).

\subsection{Limitations of the Reflection Modeling} \label{subsec:disc_limxillver}

Using \texttt{xillverTDE} to model the 1~keV line the Epoch 1 and 4 \textit{XMM-Newton}/\textit{NuSTAR} spectra requires high density ($n \gtrsim 10^{18}$ cm$^{-3}$) to produce significant oxygen K-shell features. High densities have been shown to produce significantly different reflection features compared to lower density models, especially in the soft X-ray band \citep{Garcia2016}. This is primarily due to increased temperatures in the illuminated slab due to an increase in the bremsstahlung emissivity and has been an important improvement for modeling the soft excess with relativistic reflection. The expected densities in the inner, radiation pressure dominated region of the accretion flow scale inversely with the mass of the black hole \citep[e.g.][]{Svensson1994}, and hence, high densities like those found in our fits with \texttt{xillverTDE} are expected around low mass supermassive black holes. However, with increased mass accretion rates and a geometrically thick inner accretion flow, the viscous timescales are short, implying an extremely high mass accretion rate when combined with a high density.

The illuminating spectrum in \texttt{xillverTDE} is extremely soft compared to standard coronal reflection models, making it more difficult to ionize and excite the gas around the black hole. This can be enhanced with an increase in the temperature of the gas, which occurs with increased density \citep[see][]{Garcia2016}. Thus, high density can help create the right conditions for oxygen K-shell transitions, which produces the significant soft spectral features we see with \texttt{xillverTDE}. However, in the inner accretion flow, a thick disk with $H/R \sim 1$ as shown in the right panel of Figure \ref{fig:xillverTDE} would have a viscous timescale, which scales like $(H/R)^{-2}$, on the order of hours to days, which is much shorter than the time in which the feature is observed in the spectrum ($\sim$ 1-2 years).

One possibility is to explain the high densities required by \texttt{xillverTDE} while maintaining the picture of a geometrically thick inner accretion flow is that the emission is coming from the base of the outflow. The base of the outflow is expected to be higher density than the gas in the wind itself and also explains the necessary blueshift and symmetry required by the model. Additionally, the observed density may be increased if the outflow is clumpy, and this could also potentially explain some of the rapid variability exhibited at early times when the 1~keV feature is strongest.

1~keV features have been seen in other highly accreting compact objects, including ULXs, neutron star X-ray binaries, and high-Eddington AGN, and have been successfully modeled with a variety of radiative processes, including collisionally ionized emission, photoionized emission, and soft reflection \citep[e.g.][]{Pinto2021,Ludlam2022,Xu2022}. We therefore conclude that our reflection-based modeling with \texttt{xillverTDE} is an accurate description of the data, but ultimately may not be unique. Further deep X-ray observations of super-soft, super-Eddington sources are necessary to test these soft reflection models.

\subsection{Return to Typical AGN Corona} \label{subsec:disc_corona}

1ES~1927+654 is the first AGN in which we have witnessed the destruction of the corona and a disappearing power law X-ray spectral component. As the first source to undergo such drastic X-ray evolution, extensive X-ray monitoring presented here provides a unique opportunity to study how the corona is formed and powered. 

When the power law component returns to the X-ray spectrum, the photon index is around $\Gamma \gtrsim 3$, which is much higher than what is seen in most AGN \citep[$\Gamma \approx 1.8-2.0$; e.g.][]{Nandra1994,Piconcelli2005,Winter2009,Ricci2017} and in the pre-outburst spectrum of 1ES~1927+654 \citep[$\Gamma \approx 2.4$;][]{Gallo2013}. The photon index remains roughly constant over the majority of the X-ray rise and plateau, despite rapid order of magnitude changes in the X-ray flux. The constancy of the photon index suggests a balance between heating and cooling in the corona that is steady, yet dominated by cooling more so than standard AGN corona given the abnormally high photon indices. At late times, the X-ray luminosity decreases steadily, and this is matched with a steadily decreasing photon index. This is more typical of what is seen in AGN X-ray observations, where AGN X-ray spectra appear softer when they are more luminous \citep[e.g.][]{Shemmer2006,Sobolewska2009}. 

The inclusion of \textit{XMM-Newton}/\textit{NuSTAR} observations in our analysis allows us to track not only the photon index evolution, but also the evolution of the cutoff energy of the corona, a property which was inaccessible with \textit{NICER} given its low effective area at energies $\gtrsim 2$~keV and the softness of the spectrum of 1ES~1927+654. The cutoff energy of the X-ray spectrum of 1ES~1927+654 measured with \textit{XMM-Newton}/\textit{NuSTAR} during the X-ray rise and plateau at peak luminosity is around $E_\mathrm{cut} \approx 3-8$~keV, which is extremely low compared to typical AGN values \citep[$E_\mathrm{cut} \approx 200$~keV; e.g.][]{Ricci2017}. Previous work has suggested that sources accreting at high Eddington ratios should have lower temperature corona due to increased Compton cooling from increased seed photon flux at high mass accretion rates  \citep[e.g.][]{Pounds1995}. Indeed, this trend has been observed in large surveys of AGN \citep[e.g.][]{Ricci2018}, and many highly accreting sources have been shown to have steep X-ray spectra and low temperature corona \citep[$kT_e \approx 20$~keV; e.g.][]{Kara2017}. However, a cutoff energy as low as measured in 1ES~1927+654 is unheard of, suggesting that we are seeing the system in a unique evolutionary phase.

In addition, the corona temperature in 1ES~1927+654 exhibits a cooler-when-brighter behavior, as the temperature increases as the luminosity drops in the latter half of the observations, similar to the behavior observed in two other highly accreting systems \citep[e.g. Ark 564, Swift J2127.4+5654;][]{Barua2020,Kang2021}. However, other systems have recently been shown to exhibit the opposite behavior, with the temperature of the corona increasing as the sources brighten, despite an overall softer-when-brighter behavior observed in $\Gamma$ \citep[e.g.][]{Keek2016,Zhang2018}. This hotter-when-brighter coronal behavior has been suggested to be the result of changes in the coronal geometry, potentially related to an inflated corona during the X-ray bright phases \citep[e.g.][]{Wu2020}. The difference in the trend of coronal temperature with brightness has been suggested to be dependent on the nature of the corona, namely whether the corona is close to the pair-dominated regime in compactness-temperature ($l-\Theta$) space \citep{Kang2021}. Based on our observations of 1ES~1927+654, the behavior of the corona as it is being recreated is more similar to those rapidly accreting systems that are not dominated by pair production. This is fitting with the idea that an increase in flux of seed photons can more effectively cool the corona and lead to a cooler temperature observed when the source is brighter. 

By January 2021, the cutoff energy is $E_\mathrm{cut} \gtrsim 15$~keV, which, although poorly constrained given degeneracies and limitations of the reflection modeling, is lower than most AGN, but comparable to some cool corona in nearby AGN. Unfortunately, no hard X-ray observations were taken prior to the optical outburst in 1ES~1927+654. Hence, determining the difference in coronal temperatures between the pre-outburst and latest X-ray observations is impossible. However, the similarity of the photon indices and spectral shape suggests that the corona is back to near its pre-outburst state. Further X-ray observations with deeper hard X-ray observations will be necessary to track the final state of the corona in 1ES~1927+654, but current data suggests that the corona is approaching more typical temperatures of AGN corona. 

This rapid return to an AGN-like spectrum suggests that the timescale for which the corona can be reheated is rather fast ($\sim$3-4 years). Due to the lack of X-ray observations between May 2011 and June 2018, we cannot constrain an exact timescale for recreation of the corona, but the optical/UV transient beginning in December 2017 suggests a catastrophic event that destroyed (or rapidly cooled) the corona and a timescale for corona reformation (or reheating) on the order of 3-4 years.

\subsection{Comparison to Other Nuclear Transients} \label{subsec:disc_compare}

1ES~1927+654 is a unique source with X-ray properties that we have never observed before. However, there are some properties of 1ES~1927+654 that resemble a number of other nuclear transients like CLAGN, TDEs, and QPEs. Below we discuss some comparisons to these sources and highlight the differences that make 1ES~1927+654 a unique rapid accretion event.

\subsubsection{CLAGN and TDEs}

1ES~1927+654 clearly satisfied the requirements for undergoing a so-called ``changing-look" event, with the formation of broad optical emission lines \citep{Trakhtenbrot2019}. Despite this clear association with CLAGN at optical wavelengths, the X-ray observations clearly suggest some catastrophic change to the accretion flow that is not witnessed in such extreme measures in other CLAGN \citep[e.g.][]{Parker2016,Parker2019,Wang2020a,Guolo2021}. \citet{Ricci2020} suggested that the first half of the X-ray observations of 1ES~1927+654 was consistent with a TDE in an existing AGN disk based on the $t^{-5/3}$ dependence of the optical/UV flare and the extremely soft, nearly thermal X-ray spectrum. Similar claims of TDEs in existing AGN disks have been made recently in a handful of other CLAGN \citep[e.g.][]{Merloni2015,Blanchard2017,Liu2020}, although none show as dramatic an evolution in their X-ray spectra as 1ES~1927+654. One interesting analogy is to the transient PS16dtm, which \citet{Blanchard2017} suggested could be the result of an edge-on TDE that obscures the X-ray emitting regions of a relatively face-on pre-existing accretion disk in an AGN. If the outburst in 1ES~1927+654 is also the result of a TDE in a pre-existing AGN disk, then the stark contrast in X-ray emission with PS16dtm could be the result of an optimal viewing angle to the inner accretion flow in 1ES~1927+654 (although it is possible that the lack of X-ray emission observed from PS16dtm shortly after the transient event is a result of a similar dip in X-ray flux as seen in the early observations of 1ES~1927+654). In addition to the spectral features and the optical/UV light curve, the peculiar disappearance of the power law component of the X-ray spectrum in 1ES~1927+654 is potentially consistent with recently simulations of TDEs in pre-existing AGN disks that have shown that the inner accretion flow is rapidly depleted and would hence cut off the energy supply to the corona \citep{Chan2019}. 

It is possible that the extreme changing-look event in 1ES~1927+654 was not driven by a TDE. Others have suggested that possibly a magnetic flux inversion occurred in the accretion disk of 1ES~1927+654, whereby an advection event allowed magnetic flux with opposite polarity to propagate inward in the disk \citep{Scepi2021}. The arguments supporting this idea include a delayed X-ray dip relative to the UV peak, a corresponding dip in the radio, and the return to the pre-outburst state following this event \citep{Scepi2021,Laha2022}. This model is thus in line with the interpretation put forth in this work if the anomalous flux inversion event creates a geometrically thick inner accretion flow. We would also expect a geometrically thick inner accretion flow if a TDE occurred in an existing AGN disk, and likewise we would expect that as in the magnetic flux inversion model, the system would return to its pre-outburst state after a TDE as well. The timescale for the return to a pre-outburst state would be set by a rather fast viscous timescale in the TDE case, given the high mass accretion rate and the suspected large scale height of the inner disk. We note that both the TDE and magnetic flux inversion models require a change in mass accretion rate and the destruction and recreation of the standard X-ray corona in 1ES~1927+654, although with different mechanisms triggering these changes.

\subsubsection{Quasi-Periodic Eruptions}

Additionally, the early X-ray observations of 1ES~1927+654 during the time when the light curve was rapidly variable resemble a newly discovered class of X-ray transients called quasi-periodic eruptions \citep[QPEs;][]{Miniutti2019,Giustini2020,Arcodia2021,Chakraborty2021}. QPEs were only first discovered in 2019, and currently only five QPEs are known. Hence, the mechanisms driving QPEs are currently not well understood, and there are many theories to explain their strange behavior, including interacting extreme mass ratio inspirals \citep[EMRIs; e.g.][]{King2020,Metzger2022}, self-lensing from binary black holes \citep{Ingram2021}, star-disk collisions \citep{Xian2021}, and tearing of warped accretion disks \citep[e.g.][]{Raj2021}. 

The similarities between 1ES~1927+654 and QPEs include the super-soft X-ray spectra, rapid variability on short timescales, and harder-when-brighter behavior. Phase-resolved spectral analysis of QPEs revealed that their rapid variability is likely an intrinsic change in the accretion properties and not simply due to changes in obscuration. Similarly in 1ES~1927+654, obscuration does not seem to be responsible for the rapid variability in Phase 1, as the column density of the neutral absorption in the host galaxy is always quite low ($N_H \lesssim 10^{21}$ cm$^{-2}$). However, the early X-ray light curves of 1ES~1927+654 do not exactly resemble the variable nature of QPEs, which exhibit short-duration peaks followed by periods of near quiescence with extremely low variability. In addition, the high flux spectra in QPEs often require a single additional component to account for the difference in their spectra during the flares, which due to the similarity of this component to the typical AGN soft excess lead \citet{Miniutti2019} to argue that the QPE flares could be indicative of the formation of the soft excess in AGN. This is not the case for 1ES~1927+654, which we tested by using rate-resolved spectroscopy during the highly variable \textit{XMM-Newton} observations (Epochs 2 and 3). Instead, 1ES~1927+654 required an overall change in the spectral flux and shape, suggesting an intrinsic change in the mass accretion rate. Although 1ES~1927+654 and QPEs are not exactly the same in nature, their similarities are intriguing and motivate further follow-up of super-soft transient X-ray sources.


\section{Conclusions} \label{sec:conclusion}

1ES~1927+654 is one of the most peculiar X-ray transients discovered to date. In this paper, we have shown the X-ray spectral evolution beginning shortly after the optical/UV outburst in 2018 to June 2021 using extensive observations with \textit{NICER}, \textit{XMM-Newton}, and \textit{NuSTAR}. Our main findings are summarized below:
\begin{enumerate}[noitemsep,leftmargin=*]
    \item Over the course of the past three years, the X-ray spectrum has transitioned from extremely soft and nearly thermal in the first observations to back to its pre-outburst state by January 2021. The late X-ray spectra are fit well by a significant power law component and soft excess, as in the pre-outburst spectrum. 
    \item We identify a unique three-phase evolution during the three year evolution (see Figure \ref{fig:lumin}). During Phase 1, the source exhibits rapid variability and deviates significantly from standard AGN behavior, including exhibiting harder-when-brighter behavior, a weak power law component, and a broad 1~keV line. Then, in Phase 2, 1ES~1927+654 enters a more stable super-Eddington phase, where the variability drops and the X-ray luminosity plateaus near the Eddington limit for a $10^6 \, M_\odot$ black hole. Finally, in Phase 3, 1ES~1927+654 begins to return to its pre-outburst state, as the X-ray luminosity declines in a steady manner and the spectrum hardens.
    \item We track the evolution of the power law component and find that the photon index remains relatively constant despite extreme luminosity changes on short timescales. When the X-ray luminosity starts to drop, the photon index begins to decrease and the spectrum hardens, following the typical softer-when-brighter behavior observed in AGN. 
    \item The temperature of the blackbody undergoes significant evolution over the observing period, unlike what is commonly witnessed in TDEs. We show that the blackbody deviates from the expected $L \propto T^4$ for a standard thin accretion disk during the first $\sim 1$ year of observations (i.e. Phase 1 in Figure \ref{fig:lumin}), which implies that the inner accretion flow is super-Eddington. Likewise, at late times, the blackbody has a roughly constant temperature, and this lack of temperature evolution with luminosity is consistent with observations of the soft excess in AGN.
    \item We study the evolution of the 1~keV line that was a prominent feature in the early X-ray spectra. Using the \textit{NICER} observations, we find that the equivalent width of the 1~keV feature decreases as the source approaches peak luminosity, suggesting that either the line has disappeared or has broadened significantly into a component indistinguishable from the continuum.
    \item We apply a new reflection model from a single-temperature blackbody ionizing spectrum, \texttt{xillverTDE}, to explain the 1~keV feature and find that it can fit the first \textit{XMM-Newton} spectrum well if the emission is blueshifted. We propose that this could be the result of reflected emission off of an optically thick outflow from a geometrically thick, super-Eddington inner accretion flow (see right panel of Figure \ref{fig:xillverTDE}).
    \item The \texttt{xilverTDE} model can also fit additional curvature in the peak X-ray luminosity spectrum (see Figure \ref{fig:ep4}). We also show that the standard relativistic reflection models can fit the soft excess in the latest \textit{XMM-Newton}/\textit{NuSTAR} observation, without the need for \texttt{xillverTDE} once the source has returned to its pre-outburst state.
\end{enumerate} 

Extensive X-ray monitoring of the extreme nuclear transient in 1ES~1927+654 has provided unique insight into the nature of the corona in AGN, changing-look AGN, and super-Eddington accretors. Thanks to many dedicated optical transient facilities, the number of intriguing transients is increasing significantly and can be expected to continue to increase in the coming years. Prompt X-ray follow-up of TDE-like optical light curves or extreme optical flares in existing AGN will provide valuable information about the nature of such transients and allow for more detailed studies of the population of nuclear transients. These sources are crucial probes of the inner accretion flow, disk instabilities, AGN corona formation, and many more fascinating open areas of research in accretion physics.


\medskip
\noindent We thank the anonymous referee for helpful comments that improved this manuscript. MM thanks Ruancun Li, Christos Panagiotou, and Brenna Mockler for their useful comments and discussions. EK thanks Daniel Proga and John Raymond for insightful discussions. MM and EK acknowledge support from NASA Grant 80NSSC21K0661. MM, EK, BT, and IA acknowledge support from the MISTI Global Seed Funds and the MIT-Israel Zuckerman STEM Fund. CR acknowledges support from the Fondecyt Iniciacion grant 11190831 and ANID BASAL project FB210003. RR acknowledges support from NASA Grant 80NSSC19K1287. IA is a CIFAR Azrieli Global Scholar in the Gravity and the Extreme Universe Program and acknowledges support from that program, from the European Research Council (ERC) under the European Union’s Horizon 2020 research and innovation program (grant agreement number 852097), from the Israel Science Foundation (grant number 2752/19), from the United States - Israel Binational Science Foundation (BSF), and from the Israeli Council for Higher Education Alon Fellowship.

\facilities{\textit{XMM-Newton}, \textit{NuSTAR}, \textit{NICER}}

\software{XSPEC \citep{Arnaud1996},  
Astropy \citep{AstropyCollaboration2013, AstropyCollaboration2018},
\texttt{xillver} \citep{Garcia2010,Garcia2013},
\texttt{relxill} \citep[v1.4.0;][]{Garcia2014,Dauser2014}}


\bibliography{ASASSN-18el_biblio.bib}


\appendix

\section{Representative Spectra from Each Evolutionary Phase} \label{sec:app1}

\begin{figure*}[h!]
    \centering
    \includegraphics[width=17.5cm]{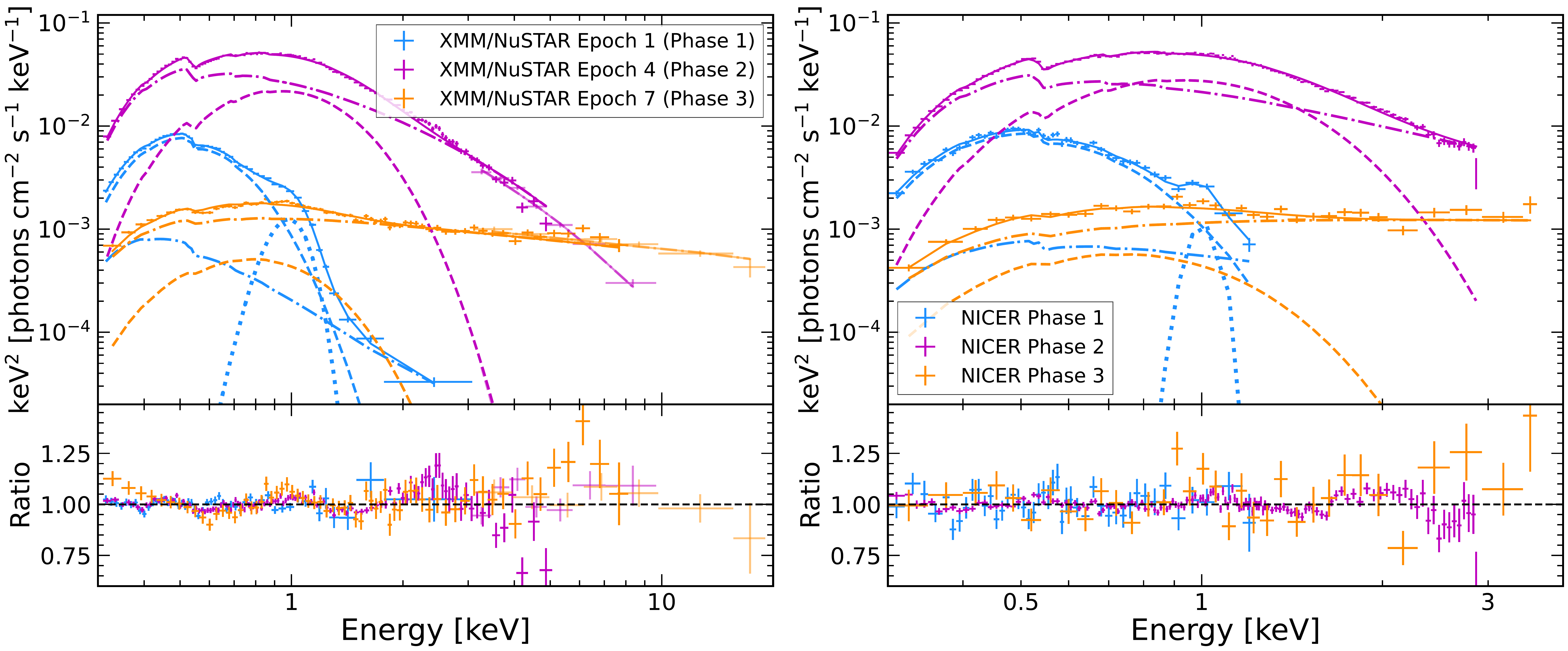}
    \caption{Representative spectra for each evolutionary phase, as designated in Figure \ref{fig:lumin}. Solid lines denoted the total model, dashed lines are the blackbody component, dot-dashed lines are the power law component, and dotted lines are the Gaussian component in each panel. \textit{Left:} The top panel shows the unfolded spectra with phenomenological models described in Section \ref{subsec:phenom_model} for \textit{XMM-Newton}/\textit{NuSTAR} observations from Epochs 1, 4, and 7. The bottom panel shows the resulting ratio between the data and the model. For visual clarity, the data has been rebinned and only data from the FPMA instrument on \textit{NuSTAR} is shown. \textit{Right:} The top panel shows the unfolded spectra with phenomenological models described in Section \ref{subsec:phenom_model} for \textit{NICER} observations taken in each evolutionary phase. The corresponding \textit{NICER} ObsIDs shown here for each phase are: Phase 1--1200190114, Phase 2--2200190331, Phase 3--3200190433. These observations were taken close to the Epoch 1, 4, and 7 observations with \textit{XMM-Newton}/\textit{NuSTAR}. The data have been rebinned for visual clarity.}
    \label{fig:phases_spectra}
\end{figure*}

In Figure \ref{fig:phases_spectra}, we show a representative spectrum from each evolutionary phase highlighted in Figure \ref{fig:lumin}, fit with the corresponding phenomenological model detailed in Section \ref{subsec:phenom_model}. The left panel of Figure \ref{fig:phases_spectra} shows the simultaneous \textit{XMM-Newton}/\textit{NuSTAR} data from Epochs 1, 4, and 7, which are used in physically-motivated reflection modeling in Section \ref{sec:physmod}. The phenomenological modeling can capture most of the significant features in these spectra, although there are some more minor residuals that we discuss in Section \ref{sec:physmod}. The right panel of Figure \ref{fig:phases_spectra} shows three representative spectra from \textit{NICER} during each of the three evolutionary phases, again fit with the phenomenological models described in Section \ref{subsec:phenom_model}. These three spectra were chosen to be taken around the same time as the simultaneous \textit{XMM-Newton}/\textit{NuSTAR} observations shown in the left panel. A very similar evolution in seen relative to the \textit{XMM-Newton}/\textit{NuSTAR} spectra. However, the \textit{NICER} spectra become background dominated at much lower energies, hence why we cannot constrain the cutoff energy of the power law component with the \textit{NICER} spectra.

\end{document}